\documentclass[11pt]{article}
\usepackage{amsmath}
\usepackage{amssymb,amsthm}
\usepackage{hyperref}
\usepackage[all]{xy}
\usepackage[english]{babel}
\usepackage[textwidth=19cm,textheight=24cm]{geometry}
\usepackage{mathrsfs}

\renewcommand{\d}{\mathrm{d}}

\newcommand{\bw}{\boldsymbol{w}}

\newtheorem{pro}{Proposition}
\newtheorem{lemma}{Lemma}
\newtheorem{definition}{Definition}
\newtheorem{theorem}{Theorem}

\newcommand{\gl}{M_N(\C)}

\newcommand{\I}{\mathbb{I}}
\newcommand{\bS}{\mathbb{S}}
\renewcommand{\d}{\mathrm{d}}

\newcommand{\bt}{\boldsymbol{t}}
\newcommand{\bs}{\boldsymbol{s}}
\newcommand{\bn}{{\boldsymbol{n}}}

\newcommand{\Exp}[1]{\operatorname{e}^{#1}}

\newcommand{\s}{\operatorname{sl}}
\newcommand{\sg}{\operatorname{sg}}

\newcommand{\T}{\mathfrak t}
\newcommand{\g}{\mathfrak{g}}

\newcommand{\Cc}{\mathcal{C}}

\newcommand{\m}{\mathcal{M}}

\renewcommand{\L}{\mathscr L}
\newcommand{\K}{\mathscr K}
\newcommand{\W}{\mathscr W}
\newcommand{\M}{\mathscr M}

\newcommand{\Pp}{\mathscr P}
\newcommand{\Ss}{\mathscr S}

\newcommand{\Z}{\mathbb Z}
\newcommand{\C}{\mathbb C}
\newcommand{\N}{\mathbb N}

\DeclareMathAlphabet{\mathpzc}{OT1}{pzc}{m}{it}
\newcommand{\Tp}{\Large\mathpzc T}

\newcommand{\tb}{\Large\mathpzc b}
\begin{document}

\title{The multicomponent 2D Toda hierarchy:\\ dispersionless limit}

\author{Manuel Ma\~{n}as and  Luis Mart\'{\i}nez Alonso\\
Departamento de F\'{\i}sica Te\'{o}rica II, Universidad Complutense\\ 28040-Madrid, Spain\\
emails: manuel.manas@fis.ucm.es, luism@fis.ucm.es} \maketitle

\abstract{
The factorization problem of the multi-component 2D Toda hierarchy is used to analyze the dispersionless limit of this
hierarchy. A dispersive version of the Whitham hierarchy defined in terms of scalar Lax and Orlov--Schulman operators
is introduced and the corresponding additional symmetries and string equations are discussed.
Then, it is shown how KP and Toda pictures of the
dispersionless Whitham hierarchy emerge in the dispersionless limit. Moreover,
the additional symmetries and string equations for the dispersive Whitham hierarchy are studied in this limit.
}


\section {Introduction}

In \cite{nuestro} the theory  of the multi-component Toda hierarchy
\cite{ueno-takasaki} was analyzed from the point of view of a
factorization problem
\begin{gather}
  \label{facW0}
 g=W^{-1}\,\bar W
\end{gather}
in an infinite-dimensional group and a natural formulation of the
additional symmetries and the string equations of the hierarchy was
given. In the present work we use this formulation to study the
dispersionless limit of the solutions of \eqref{facW0}. As it is
known in the theory of random matrix models \cite{gin}-\cite{bon},
the study the large $N$ limit can be performed
 in terms of the dispersionless limit of  the string equations satisfied by the solution of the underlying integrable
 system. Notice that in recent years the formalism of  string equations \cite{takasaki string}
 for dispersionless integrable systems \cite{takasaki-takebe} has been much developed  \cite{string}. Our present work is
motivated by the applications of multi-component integrable
hierarchies \cite {ueno-takasaki,KP} to the study of the large $N$
limit of the two-matrix model \cite{b}-\cite{adler0}, as well as
models of random matrices with external source and non-intersecting
Brownian motions \cite{adler2}-\cite{d}. A common feature of these
models is that they have an associated family of multiple orthogonal
polynomials which is in turn characterized by a matrix
Riemann-Hilbert (MRH) problem which is a basic ingredient to analyze
the large $N$ limit \cite{d}-\cite{dei2}. On the other hand, MRH
problems also provide solutions of reductions  of multi-component
integrable hierarchies of KP or Toda type. These reductions
correspond  to solutions of factorization problems \eqref{facW0}
constrained by certain types of string equations.

In our analysis we introduce matrix wave functions and  scalar Lax
and Orlov--Schulman operators \cite{orlov} associated to the
solutions of \eqref{facW0}. We prove that the rows  of the matrix
wave functions satisfy auxiliary linear systems involving the scalar
Lax operators, which constitute the dispersive versions of the genus
zero dispersionless Whitham hierarchies \cite{krichever}. In order
to study the dispersionless limit,  we assume the Takasaki--Takebe
quasi-classical ansatz \cite{misgam,takasaki-takebe-ultimo} for the
rows of the matrix wave functions. Thus, we prove that in the
dispersionless limit the auxiliary linear systems reduce to systems
of Hamilton--Jacobi equations that are shown to be equivalent to the
dispersionless Whitham hierarchies. In particular, two natural
pictures
 (KP and Toda types) of the dispersionless Whitham hierarchies emerge in our analysis. An important advantage of our
approach is that it yields a natural method for characterizing
string equations and additional symmetries in the dispersionless limit. In particular,
we characterize the dispersive analogues of the soluble string equations
discussed in \cite{mano1}.

 The layout of the paper is as follows. In \S 1.1 we present a summary of the relevant parts of
 \cite{nuestro} needed in the subsequent analysis. Then, in \S 2 we discuss the dispersive Whitham hierarchies.
  We introduce
 a set of scalar
 Lax and Orlov--Schulman operators, and vector wave functions to deduce the corresponding
auxiliary linear systems, as well as additional symmetries and string equations of dispersive type.
 Finally, in \S 3 we discuss the aforementioned dispersionless limits. We find the Hamilton--Jacobi type equations,
 and then derive  the KP and Toda pictures of the dispersionless Whitham hierarchy.
  We conclude the paper by considering the dispersionless counterparts  of dispersive string equations.

\subsection{Reminder}
 As in our previous work \cite{nuestro} we  only consider formal series expansions in the Lie group theoretic set up without
 any assumption on their convergency.  Let us remind some notations and results from \cite{nuestro}. Given
Lie algebras $\g_1\subset \g_2$, and $X,Y\in\g_2$ then
 $ X=Y+\g_1 $ means $X-Y\in\g_1$. For any Lie groups $G_1\subset G_2$ and $a,b\in G_2$ then
$  a=G_2\cdot b $ stands for $a\cdot b^{-1}\in G_2$.

Let $\gl$ denote the associative algebra  of complex $N\times N$
complex matrices we will consider the linear space of sequences
$f:\Z\rightarrow M_N(\C)$.
 The shift operator $\Lambda$ acts on these sequences as
$(\Lambda f)(n):=f(n+1)$. A sequence $X:\Z\to \gl$ acts by left
multiplication in this space of sequences, and therefore we may
consider operators of the type $X\Lambda^j$, $(
X\Lambda^j)(f)(n):=X(n)\cdot f(n+j)$.

Moreover, defining the product
$(X(n)\Lambda^i)\cdot(Y(n)\Lambda^j):=X(n)Y(n+i)\Lambda^{i+j}$
and extending it linearly we have that the set $\g$ of Laurent
series in $\Lambda$ is an associative algebra, which under the standard commutator is a
Lie algebra.

This Lie algebra has the following important splitting
\begin{gather}\label{splitting}
\g=\g_+\dotplus\g_-,
\end{gather}
where
\begin{align*}
  \g_+&=\Big\{\sum_{j\geq 0}X_j(n)\Lambda^j,\quad X_j(n)\in\gl\Big\},&
  \g_-&=\Big\{\sum_{j< 0}X_j(n)\Lambda^j,\quad X_j(n)\in\gl\Big\},
\end{align*}
are Lie subalgebras of $\g$ with trivial intersection.

The group of linear invertible elements in $\g$ will be denoted by
$G$ and has $\g$ as its Lie algebra, then the splitting
\eqref{splitting} leads us to consider the following factorization of
$g\in G$
\begin{gather}\label{fac1}
g=g_-^{-1}\cdot g_+, \quad g_\pm\in G_\pm
\end{gather}
where $G_\pm$ have $\g_\pm$ as their Lie algebras. Explicitly, $G_+$
is the set of invertible linear operators  of the
form $\sum_{j\geq 0}g_j(n)\Lambda^j$; while $G_-$ is the set of
invertible linear operators of the form
$1+\sum_{j<0}g_j(n)\Lambda^j$.

 Now we
introduce two sets of indexes, $\mathbb S=\{1,\dots,N\}$ and
$\bar{\mathbb S}=\{\bar 1,\dots,\bar N\}$, of the same cardinality
$N$. In what follows we will use letters $k,l$  and $\bar k,\bar l$ to
denote elements in  $\mathbb S$ and $\bar{\mathbb S}$,
 respectively. Furthermore, we will use letters $a,b,c$ to denote elements in
 $\mathcal S:=\mathbb S\cup\bar{\mathbb
S}$.

We define the following operators $ W_0,\bar  W_0\in G$
\begin{align}
 \label{def:E}  W_0&:=\sum_{k=1}^NE_{kk}\Lambda^{s_k}\Exp{\sum_{j=0}^\infty
 t_{jk}\Lambda^{j}}, \\
\label{def:barE}   \bar W_0&:=\sum_{k=1}^NE_{kk}\Lambda^{- s_{\bar k}}\Exp{\sum_{j=1}^\infty
   t_{j\bar k}\Lambda^{-j}}
\end{align}
where $s_a\in \Z,\, t_{ja}\in \C$ are deformation parameters, that in the sequel
will play the role of discrete and continuous times, respectively.
Given an element $g\in G$ and a set of deformation
parameters $\bs=(s_a)_{a\in\mathcal S},\bt=(t_{ja})_{a\in\mathcal
S,j\mathbb\in \N}$ we will consider the factorization problem
\begin{gather}
  \label{factorization}
  S(\bs,\bt)\cdot W_0\cdot g=\bar S(\bs,\bt)\cdot\bar W_0,\quad S\in G_-\text{ and } \bar S\in G_+,
\end{gather}
and will confine ourselves to the \emph{zero charge sector} $
|\bs|:=\sum_{a\in \mathcal S}s_a=0$.
 We   define the dressing or Sato operators $W,\bar W$ as follows
\begin{align}
\label{def:baker}W&:=S\cdot W_0,& \bar W&:=\bar S\cdot \bar  W_0,
\end{align}
so that the factorization problem in $G$ reads
\begin{gather}
  \label{facW}
  W\cdot g=\bar W
\end{gather}
Observe that  $S,\bar S$ have expansions of the form
\begin{gather}
\label{expansion-S}
\begin{aligned}
S&=\I_n+\varphi_1(n)\Lambda^{-1}+\varphi_2(n)\Lambda^{-2}+\cdots\in G_-,\\
\bar
S&=\bar\varphi_0(n)+\bar\varphi_1(n)\Lambda+\bar\varphi_2(n)\Lambda^{2}+\cdots\in
G_+.
\end{aligned}
\end{gather}

 The Lax  operators $L,\bar L,C_{kk},\bar C_{kk}\in\g$
 are defined by
\begin{align}
\label{Lax}  L&:=W\cdot\Lambda\cdot W^{-1}, & \bar L&:=\bar W\cdot\Lambda\cdot \bar W^{-1}, \\
\label{C} C_{kk}&:=W\cdot E_{kk}\cdot W^{-1},& \bar C_{kk}&:=\bar
W\cdot E_{kk}\cdot \bar W^{-1}
\end{align}
and
have the following expansions
\begin{gather}\label{lax expansion}
\begin{aligned}
 L&=\Lambda+u_1(n)+u_2(n)\Lambda^{-1}+\cdots,&
\bar L^{-1}&=\bar u_0(n)\Lambda^{-1}+\bar u_1(n)+\bar u_2(n)\Lambda+\cdots, \\
C_{kk}&=E_{kk}+C_{kk,1}(n)\Lambda^{-1}+C_{kk,2}(n)\Lambda^{-2}+\cdots,&
\bar \Cc_{kk}&=\bar C_{kk,0}(n)+\bar C_{kk,1}(n)\Lambda+\bar
C_{kk,2}(n)\Lambda^{2}+\cdots.
\end{aligned}
\end{gather}

Now we introduce some further notation
\begin{enumerate}
\item\begin{align*}
&  \partial_{ja}:=\frac{\partial}{\partial t_{ja}},\quad \text{for
$a=\mathcal S$ and $j=1,2,\dots$}
\end{align*}
\item Given $K=(a,b)$ the basic \emph{charge preserving} shift operators
$T_{K}$  are defined as follows
\begin{equation*}
(T_{K}\,f)(s_a, s_b):=f(s_a+1, s_b-1).
\end{equation*}
\end{enumerate}

We define the Orlov--Schulman operators \cite{orlov} for the
multi-component 2D Toda hierarchy by
\begin{align}
  \label{orlov}
  M&:=Wn W^{-1},& \bar M&:=\bar Wn \bar W^{-1}.
\end{align}
One proves at once that
\begin{itemize}
\item The Orlov--Schulman operators satisfy the following commutation relations
\begin{gather}\label{algrebraic-orlov}
\begin{aligned}
{}[L,M]&=L,& [L,C_{kk}]&=0,& [\bar L,\bar M]&=\bar L,& [\bar L,\bar
C_{kk}]&=0,
\end{aligned}
\end{gather}
\item The following expansions hold
\begin{gather}\label{orlov-exp}
\begin{aligned}
  M&=\m+
  \sum_{k=1}^NC_{kk}(s_k+\sum_{j=1}^\infty j t_{jk}L^j),& \m&=n+ \g_-\\
\bar M&=\bar \m-
  \sum_{k=1}^N\bar C_{kk}( s_{\bar k}+\sum_{j=1}^\infty j  t_{j\bar k}\bar L^{-j}),&\bar \m&=n+\g_+\Lambda.
\end{aligned}
\end{gather}
\end{itemize}

\subsubsection{Additional symmetries}
Suppose that the operator $g$ in \eqref{facW} depends on an
additional parameter $\tb\in\C$. Then, the basic objects of the multi-component Toda
 hierarchy inherit a dependence on $\tb$ .
For convenience and for the time being we use the following  equivalent
factorization problem
\begin{align*}
W\cdot h=\bar W\cdot\bar h,
\end{align*}
with
\begin{align}\label{factor g}
g=h\cdot\bar h^{-1}.
\end{align}
Observe that
\begin{align}\label{external parameter continous}
\begin{aligned}
\partial_{\tb}W\cdot W^{-1}+W(\partial _{\tb}h\cdot h^{-1})W^{-1}&=\partial_{\tb} S\cdot S^{-1}+W(\partial _{\tb}h\cdot h^{-1})W^{-1}\\&=\partial_{\tb}\bar S\cdot\bar S^{-1}+\bar W(\partial _{\tb}\bar h\cdot\bar h^{-1})\bar W^{-1}=
\partial_{\tb}\bar W\cdot \bar W^{-1}+\bar W(\partial _{\tb}\bar h\cdot\bar h^{-1})\bar
W^{-1}.
\end{aligned}
\end{align}
Now, let us  suppose that $h$ and $\bar h$ satisfy
\begin{align}\label{derivada g}
  \begin{aligned}
    \partial_{\tb}h\cdot h^{-1}&=F^{(0)}=\sum_{l=1}^N F_{l}(n,\Lambda)E_{ll},  &
     \partial_{\tb}\bar h\cdot \bar h^{-1}&=\bar F^{(0)}=\sum_{l=1}^N \bar
     F_{l}(n,\Lambda)E_{ll},
  \end{aligned}
\end{align}
then from \eqref{external parameter continous}we get
\begin{align*}
   \partial_{\tb} W\cdot W ^{-1}= \partial_{\tb} S\cdot S^{-1}&=-H_-, &
   \partial_{\tb} \bar W\cdot \bar W^{-1}=\partial_{\tb} \bar S\cdot \bar S^{-1}&=H_+,  &H_\pm&\in\g_\pm.
\end{align*}
where
\begin{align}
  &H:=F-\bar F,&
  \label{fbarf}
    F&:=\sum_{l=1}^N F_{l}(M,L)C_{ll},&
     \bar F&:=\sum_{l=1}^N \bar F_{l}(\bar M,\bar L)\bar C_{ll}.
\end{align}

 Hence it follows that
\begin{pro}\label{pro:external parameters}
Given a dependence on an additional parameter $\tb$ according to
\eqref{factor g}, \eqref{derivada g} and \eqref{fbarf} then
\begin{enumerate}
  \item The dressing operators $W$ and $\bar W$ satisfy
  \begin{align*}
    \partial_{\tb} W&=-H_-\cdot W, &\partial_{\tb}\bar W&=H_+\cdot\bar W,
 \end{align*}
\item The Lax and Orlov--Schulman operators satisfy
\begin{align}\label{additional flows}
\begin{aligned}
   \partial_{\tb}  L&=[-H_-,L],& \partial_{\tb} M&=[-H_-,M],& \partial_{\tb}  C_{kk}&=-[H_-,C_{kk}],\\
  \partial_{\tb}  \bar L&=[H_+,\bar L],& \partial_{\tb}  \bar M&=[H_+,\bar M],& \partial_{\tb} \bar C_{kk}&=[H_+,\bar C_{kk}].
  \end{aligned}
\end{align}
\end{enumerate}
\end{pro}

A key observation is
\begin{pro}\label{gd}
Given operators $R, \bar R\in\g$ satisfying $ R\cdot g=\bar R$ and
such that
\begin{align}\label{nuevolemma}
\begin{aligned}
  RW_0^{-1}\in\g_-,\\
  \bar R\bar W_0^{-1}\in\g_+.
\end{aligned}
\end{align}
Then
$
  R=\bar R=0
$
\end{pro}

\subsection{Wave functions}

The wave functions of the multi-component 2D Toda hierarchy are
defined by
\begin{gather}\label{baker-fac}
\begin{aligned}
\psi&= W\cdot\chi, &
\bar\psi&=\bar W\cdot\chi.
\end{aligned}
\end{gather}
where
\[
\chi(z):=\{z^n\mathbb I_N\}_{n\in\Z},\
\]
Note that $\Lambda\chi=z\chi$. The following asymptotic expansions
are a consequence of \eqref{expansion-S}
\begin{gather}\label{baker-asymp}
\begin{aligned}
  \psi&=z^n(\I_N+\varphi_1(n)z^{-1}+\cdots)\,\psi_0(z),&\psi_0&:=\sum_{k=1}^NE_{kk}
z^{s_k} \Exp{\sum_{j=1}^\infty t_{jk}z^j},& z&\rightarrow\infty,\\
\bar\psi&=z^n(\bar\varphi_0(n)+\bar\varphi_1(n)z+\cdots)\,\bar\psi_0(z),
&\bar\psi_0&:=\sum_{k=1}^NE_{kk} z^{- s_{\bar k}}
\Exp{\sum_{j=1}^\infty  t_{j\bar k}z^{-j}},& z&\rightarrow 0.
\end{aligned}
\end{gather}

\begin{pro}\label{Lax C baker}
\begin{enumerate}
  \item Given operators of the form
 \begin{align*}
   F&:=\sum_{k=1}^NF_kC_{kk},& F_k&:=\sum_{i\geq 0,j\in \Z}F_{kij}M^iL^j,&
      \bar F&:=\sum_{k=1}^N F_{\bar k}\bar C_{kk},&  F_{\bar k}&:=\sum_{i\geq 0,j\in \Z} F_{\bar kij}\bar M^i\bar L^j,
 \end{align*}
 with complex-valued scalar coefficients, we
 have
 \begin{align}\label{zpsi}
   F(\psi)&=(\psi)\sum_{k=1}^N
   \overleftarrow{F_k\Big(z\frac{\d}{\d z},z\Big)}E_{kk},&
    \bar F(\bar\psi)&=(\bar \psi)\sum_{k=1}^N\overleftarrow{ F_{\bar k}\Big(z\frac{\d}{\d z},z\Big)}E_{kk}.
 \end{align}
where
\begin{align*}
   (\psi) \overleftarrow{F_k\Big(z\frac{\d}{\d z},z\Big)}:=&\sum_{i\geq 0,j\in \Z}F_{kij}  z^j\Big( z\frac{\d }{\d z}\Big)^i(\psi),&  (\bar\psi) \overleftarrow{ F_{\bar k}\Big(z\frac{\d}{\d z},z\Big)}:=&\sum_{i\geq 0,j\in \Z}\bar F_{kij}  z^j\Big( z\frac{\d }{\d z}\Big)^i(\bar\psi)
\end{align*}
\item Given operators
 \begin{align*}
   P&:=\sum_{k=1}^NP_kC_{kk},& P_k&:=\sum_{i\geq 0,j\in \Z}P_{kij}M^iL^j,&
    Q&:=\sum_{k=1}^NQ_kC_{kk},& Q_k&:=\sum_{i\geq 0,j\in \Z}Q_{kij}M^iL^j,\\
    \bar  P&:=\sum_{k=1}^N P_{\bar k}\bar C_{kk},&  P_{\bar k}&:=\sum_{i\geq 0,j\in \Z} P_{\bar kij}\bar M^i\bar L^j,&   \bar  Q&:=\sum_{k=1}^N  Q_{\bar k}\bar C_{kk},&   Q_ {\bar k}&:=\sum_{i\geq 0,j\in \Z}\bar Q_{\bar kij}\bar M^i\bar L^j,
 \end{align*}
 with complex-valued scalar coefficients, we
 have
 \begin{align*}
   PQ(\psi)&=\sum_{k=1}^N\bigg((\psi) \overleftarrow{P_k\Big(z\frac{\d}{\d z},z\Big)} \bigg) \overleftarrow{Q_k\Big(z\frac{\d}{\d z},z\Big)},&
     \bar P\bar Q(\bar\psi)&=\sum_{k=1}^N \bigg((\bar\psi)\overleftarrow{ P_{\bar k}\Big(z\frac{\d}{\d z},z\Big)} \bigg) \overleftarrow{  Q_{\bar k}\Big(z\frac{\d}{\d z},z\Big)}.
 \end{align*}
 \end{enumerate}
\end{pro}
\begin{proof}
\begin{enumerate}
 \item
  It is easy to find from  \eqref{Lax} and \eqref{orlov-exp} that
  \begin{align*}
  M^iL^jC_{kk}(\psi)&=Wn^i\Lambda^jE_{kk}(\chi)=W\Big(\{n^iz^{n+j}\I_N\}_{n\in\Z}\Big)E_{kk},\\
  \bar M^i\bar L^j\bar C_{kk}(\bar \psi)&=\bar Wn^i\Lambda^jE_{kk}(\chi)=\bar W\Big(\{n^iz^{n+j}\I_N\}_{n\in\Z}\Big)E_{kk}.
  \end{align*}
    Now observe that the action of $X=\sum_{j'\in \Z}X_{j'}\Lambda^{j'}$  on $\{n^iz^{n+j}\}_{n\in\Z}$ is
    \begin{align*}
      X\Big(\{n^iz^{n+j}\}_{n\in\Z}\Big)=\Big\{\sum_{j'\in \Z}X_{j'}(n)(n+j')^iz^{n+j+j'}\Big\}_{n\in\Z}
    \end{align*}
    or equivalently
    \begin{align*}
     z^j\Big( z\frac{\d }{\d z}\Big)^i(X\cdot\chi)=\bigg\{z^j\Big(z\frac{\d}{\d z}\Big)^i\Big(\sum_{j'\in \Z}X_{j'}(n)z^{n+j'}\Big)\bigg\}_{n\in\Z}.
    \end{align*}
    Thus, the formulae
\begin{align}
  \label{z}
  M^iL^jC_{kk}(\psi)&=z^j\Big(z\frac{\d}{\d z}\Big)^i(\psi)E_{kk},&
   \bar M^i\bar L^j\bar C_{kk}(\bar \psi)&=z^{j}\Big(z\frac{\d}{\d z}\Big)^i(\bar\psi)E_{kk},
\end{align}
hold.
    \item It is a consequence of the identities
      \begin{align*}
    M^{i_1}L^{j_1}M^{i_2}L^{j_2}&=M^{i_1}(M+j_1)^{i_2}L^{j_1+j_2},
    \\
     \bar  M^{i_1}\bar L^{j_1}\bar M^{i_2}\bar L^{j_2}&=\bar M^{i_1}(\bar M+j_1)^{i_2}\bar L^{j_1+j_2},
     \\
    z^{j_2}\Big(z\frac{\d}{\d z}\Big)^{i_2}  z^{j_1}\Big(z\frac{\d}{\d z}\Big)^{i_1}&=
    z^{j_1+j_2}\Big(z\frac{\d}{\d z}+j_1\Big)^{i_2}\Big(z\frac{\d}{\d
    z}\Big)^{i_1},
  \end{align*}
    for any $i_1,i_2\geq 0$ and $j_1,j_2\in\Z$.
Therefore,
\begin{align*}
   \Big(  (\psi) \overleftarrow{\Big(z\frac{\d}{\d z}\Big)^{i_1} z^{j_1}}   \Big) \overleftarrow{\Big(z\frac{\d}
   {\d z}\Big)^{i_2}z^{j_2} }=(\psi)\overleftarrow{  \Big(z\frac{\d}{\d z}\Big)^{i_1}
    \Big(z\frac{\d}{\d z}+j_1\Big)^{i_2}z^{j_1+j_2}}=
    M^{i_1}L^{j_1}M^{i_2}L^{j_2}(\psi).
\end{align*}
\end{enumerate}

\end{proof}

\section{The dispersive Whitham hierarchies} As we will see  certain families of equations of the multi-component 2D Toda hierarchy,
associated with any given row of the dressing operators, become the
Whitham hierarchies under appropriate dispersionless limits. Consequently, these
families will be referred to as the dispersive Whitham hierarchies.

For simplicity and without loss of generality,  we will work with
the first row of the dressing operators. It will be
useful to introduce the following  shift operators
\begin{align}
  \Tp_a&:=\begin{cases}
    T_{(1,a_0)},& a=1,\\
    T_{(a,1)}, &a\neq 1
  \end{cases}&
   \bar \Tp_a&:=\begin{cases}
    T_{(\bar 1,a_0)},& a=\bar 1,\\
    T_{(a,\bar 1)}, &a\neq \bar 1,
  \end{cases}
\end{align}
where for the cases $a=1$ and $a=\bar 1$, the index $a_0$ stands for
any fixed elements in $\mathcal{S}-\{1\}$ and
$\bar{\mathcal{S}}-\{\bar 1\}$, respectively. These two types of
shift operators, that we refer as bared and unbared,  lead to
two algebras of shift operators, and also to two different families
of Hamilton--Jacobi equations, see \eqref{HJ} and \eqref{barHJ}. We
also define the scalar dressing operators
\begin{align}
  \label{Koperator}
  \K_{a}& := \begin{cases}
  1+\varphi_{1,11}\Tp_1^{-1}+\varphi_{2,11}\Tp_l^{-2}+\cdots, &  a= 1\\
 \varphi_{1,1k}+\varphi_{2,1k}\Tp_k^{-1}+\cdots,&a= k\neq 1,
 \\ \bar\varphi_{0,1k}+\bar\varphi_{1,1k}\Tp_{\bar k}^{-1}+\cdots, &a= \bar k,
    \end{cases} &
\\\bar \K_a&:=  \begin{cases}
  1+\varphi_{1,11}\bar\Tp_{ 1}^{-1}+\varphi_{2,ll}\bar\Tp_{ 1}^{-2}+\cdots, &  a= 1,\\
  \varphi_{1,lk}+\varphi_{2,lk}\bar\Tp_{k}^{-1}+\cdots,&a= k\neq 1\\
   \bar\varphi_{0,1k}+\bar\varphi_{1,1k}\bar\Tp_{\bar k}^{-1}+\cdots, &a= \bar k,      \end{cases}
   \end{align}
where $\varphi_i,\bar\varphi_i$ are the matrix coefficients of \eqref{expansion-S}.

Thus, we may now introduce the associated scalar  Lax  operators
\begin{align}\label{ttlax}
\begin{aligned}
\L_a&:=
  \K_{a} \circ {\Tp}_{a}\circ  \K_{a}^{-1}=\W_a \circ {\Tp_a}\circ  \W_a^{-1}=
\begin{cases}
  \Tp_1+L_{1,0}
+L_{1,-1}\Tp_1^{-1}+\cdots,& a=1,\\
L_{a,1}  \Tp_a+L_{a,0}+L_{a,-1}\Tp_a^{-1}+\cdots, &a\neq 1
  \end{cases}
\\
 \bar\L_{a}&:=
 \bar \K_{a} \circ \bar\Tp_{a}\circ  \bar\K_{a}^{-1}=  \bar \W_{a} \circ \bar\Tp_{a}\circ  \bar\W_{a}^{-1}=\begin{cases}
  \bar\Tp_1+
\bar L_{1,0}+\bar L_{1,-1}\bar\Tp_1^{-1}+\cdots,& a=1,\\
 \bar L_{a,1}\bar\Tp_a+\bar L_{a,0}+\bar L_{a,-1}\bar\Tp_a^{-1}+\cdots, &a\neq 1,
  \end{cases}
\end{aligned}
\end{align}
where
\begin{align}\label{wave}
 \W_a&:=\K_a\circ\W_{0,a},& \W_{0,a}&:=\exp(\mathscr T_a),& \mathscr T_a&:=\sum_{j=1}^\infty t_{ja}\Tp_a^j,
\\
\bar\W_a&:=\bar
\K_a\circ\bar\W_{0,a},&\bar\W_{0,a}&:=\exp(\bar{\mathscr T}_a), &\bar{\mathscr T}_a&:=
 \sum_{j=1}^\infty t_{ja}\bar\Tp_a^j.
\end{align}

Similarly, we define  the corresponding  scalar Orlov--Schulman
operators
 by
\begin{align}\label{orlov whitham}
  \M_a&:=n-\nu_a+\sg (a)\W_a\circ s_a\circ \W_a^{-1},&\bar \M_a&:=n-\nu_{a}+\sg (a)\bar \W_a\circ s_a\circ \bar\W_a^{-1},&
\end{align}
where
\begin{align*}
 \sg(a)&:=\begin{cases}
    1, &a\in\mathbb S,\\
   -1, & a\in\bar{\mathbb S},
  \end{cases}&
  \nu_{a}&:= \begin{cases}
    1,& a\in\mathbb S-\{1\},\\
    0, & a\not\in \mathbb S-\{1\}.
  \end{cases}
\end{align*}
  From the identities
  \begin{align*}
    [\Tp_a,\sg(a)s_a]&=\sg(a)\Tp_a,& [\bar\Tp_a,\sg(a)s_a]&=\sg(a)\bar\Tp_a,
  \end{align*}
it follows that
    \begin{align*}
    [\L_a,\M_a]&=\sg(a)\L_a,& [\bar \L_a,\bar\M_a]&=\sg(a)\bar\L_a,
  \end{align*}
\begin{pro}
  The  Orlov--Schulman   operators  satisfy
   \begin{align}\label{orlov whitham expressions}
  \M_a&=n-\nu_{a}+\sg(a)\Big(s_{a}+
   \sum_{j=1}^\infty jt_{j a}\L_{a}^j+ \sum_{i=1}^\infty m_{ai}\Tp_a^{-i}\Big),\\
   \label{bar orlov whitham expressions}
      \bar \M_a&:=  n-\nu_{a}+ \sg(a)\Big(s_a+ \sum_{j=1}^\infty jt_{ja}\bar \L_a^j+
      \sum_{i=1}^\infty\bar m_{ai}\bar\Tp_a^{-i}\Big).
     \end{align}
\end{pro}
\begin{proof}
  These formulae follow from
  \begin{align*}
    \W_{0,a} s_a \W_{0,a}^{-1}&=s_a+[\mathscr T_a,s_a]=
      s_a+\sum_{j=1}^\infty j \Tp_a^j,&
           \bar \W_{0,a} s_a \bar\W_{0,a}^{-1}&=s_a+[\bar {\mathscr T}_a,s_a]=
      s_a+\sum_{j=1}^\infty j \bar\Tp_a^j,
  \end{align*}
  and the fact that there are expansions of the form
  \begin{align*}
    \K_a s_a\K_a^{-1}&= s_a+\sum_{i=1}^\infty m_{ai}\Tp_a^{-i},&  \bar\K_a s_a\bar\K_a^{-1}&=
       s_a+\sum_{i=1}^\infty \bar m_{ai}\bar\Tp_a^{-i}.
   \end{align*}
\end{proof}

We further introduce the vector wave functions
 \begin{align}\label{PSI}
\Psi_a&:=\begin{cases}
  \psi_{1k},& a=k,\\
  \bar \psi_{1k}, & a=\bar k,
\end{cases}
\end{align}

\begin{pro}\label{pro: shift orlov}
We have the identities
  \begin{align}\label{bnota1}
[ F_a(\M_a,\L_a)]( \Psi_{a})=[ F_a(\bar {\M}_a,\bar{\L}_a)](
\Psi_{a})&=(\Psi_{a})\overleftarrow{F_a\Big(z\frac{\d}{\d z},z^{\sg
a}\Big)}=\begin{cases}
 E_{11}F_k(M,L)C_{kk}(\psi),& a=k,\\
 E_{11} F_{\bar k}(\bar M, \bar L^{-1})\bar C_{kk}(\bar\psi),& a=\bar k.
\end{cases}
\end{align}
\end{pro}
\begin{proof}
 From the definitions \eqref{def:E},\eqref{def:barE} and \eqref{def:baker}
  \begin{align*}
    W_{1k}n^i\Lambda^j=S_{1k}W_{0,kk}n^i\Lambda^j=S_{1k}\,(W_{0,kk}nW_{0,kk}^{-1})^i\,
    \Lambda^j W_{0,kk}=S_{1k}\,(n+s_k+\sum_{j'=1}^\infty j'
    t_{j'k}\Lambda^{j'})^i\,
    \Lambda^j W_{0,kk},\\
 \bar   W_{1k}n^i\Lambda^j=\bar S_{1k}\bar W_{0,kk}n^i\Lambda^j=\bar S_{1k}(\bar W_{0,kk}n\bar W_{0,kk}^{-1})^i
    \Lambda^j \bar W_{0,kk}=\bar S_{1k}(n-s_{\bar k}-\sum_{j'=1}^\infty j' t_{j'\bar k}\Lambda^{-j'})^i
    \Lambda^j \bar W_{0,kk},
  \end{align*}
 Now, observe that
  \begin{align*}
  \Lambda^{-1}(n+s_k)W_{0,kk}&=\Tp_k^{-1}((n+s_k)W_{0,kk})=\bar\Tp_k^{-1}((n+s_k)W_{0,kk}),\\
   \Lambda(n-s_{\bar k})\bar W_{0,kk}&=\Tp_{\bar k}^{-1}((n-s_{\bar k})\bar W_{0,kk})=\bar\Tp_{\bar k}^{-1}
   ((n-s_{\bar k})\bar W_{0,kk}),
    \end{align*}
together with Proposition \ref{Lax C baker} imply the  result.
  \end{proof}

\subsection{Auxiliary linear systems}

Our next analysis uses the following complex algebras
\begin{align}
  \T_a&:=\Big\{\sum_{j\in\Z} c_j \Tp_a^j\Big\},&   \bar\T_{a}&:=\Big\{\sum_{j\in\Z}  c_j \bar \Tp_a^j\Big\},
\end{align}
 and their subalgebras
\begin{align}\label{splitt}
&\left\{  \begin{aligned}
    \T_{a,+}=\T_{a,>}&:=\Big\{\sum_{j>0} c_j \Tp_a^j\Big\}& \T_{a,\leq}&:=\Big\{\sum_{j\leq 0} c_j \Tp_a^j\Big\},&a\neq 1\\  \T_{1,+}=\T_{1,\geq}&:=\Big\{\sum_{j\geq 0} c_j \Tp_1^j\Big\}& \T_{1,<}&:=\Big\{\sum_{j< 0} c_j \Tp_1^j\Big\}
  \end{aligned}\right.\\
& \left\{  \begin{aligned}
   \bar\T_{a,+}=\bar \T_{a,>}&:=\Big\{\sum_{j>0} c_j (\bar\Tp_a^j-1)\Big\}, \quad & \bar  \T_{a,<}&:=\Big\{\sum_{j< 0} c_j(\bar \Tp_a^j-1)\Big\},&a\neq 1, \bar 1\\
      \bar\T_{1,+}=   \bar \T_{1,\geq}&:=\Big\{\sum_{j\geq 0} c_j \bar\Tp_1^j\Big\}, \quad & \bar  \T_{1,<}&:=\Big\{\sum_{j< 0} c_j\bar \Tp_1^j\Big\},&\\
        \bar\T_{\bar 1,+}= \bar \T_{\bar 1,>}&:=\Big\{\sum_{j> 0} c_j (\bar\Tp_{\bar 1}^j-1)\Big\}& \bar\T_{\bar 1,<}&:=\Big\{\sum_{j<0} c_j (\bar\Tp_{\bar 1}^j-1)\Big\}, &a'\neq 1,\\
            \bar\T_{\bar 1,+}= \bar \T_{\bar 1,>}&:=\Big\{\sum_{j> 0} c_j \bar\Tp_{\bar 1}^j\Big\}& \bar\T_{\bar 1,\leq}&:=\Big\{\sum_{j\leq0} c_j \bar\Tp_{\bar 1}^j\Big\}, &a'= 1.
  \end{aligned}\right.\label{splitt1}
  \end{align}
   We will
  denote by  ($T_{a,+},T_{a,<},T_{a,>},T_{a,\leq},T_{a,\geq}$) the projections of an operator
  $T_a$ induced by the  corresponding splittings. 

The following important result links the operators
$(\M_k,\L_k)$ with the operators $(M,L)$
. Here the splittings for
each shift algebra $\T_a$ or $\bar\T_a$ are those indicated by
\eqref{splitt} and \eqref{splitt1},
\begin{pro}\label{tec.pro}
  The following relations hold
    \begin{align}
 \label{F1} & \left\{   \begin{aligned}
        F(\M_k,\L_k)_+(E_{11}W)&=     F(\bar\M_k,\bar\L_k)_+(E_{11}W)=E_{11}(F(M,L)C_{kk})_+W,\\
           F(\M_k,\L_k)_+(E_{11}\bar W)&=F(\bar\M_k,\bar\L_k)_+(E_{11}\bar W)=E_{11}(F(M,L)C_{kk})_+\bar
           W,
      \end{aligned}
      \right.&
      \\
      \label{F3}  & \left\{   \begin{aligned}
       F(\M_{\bar k},\L_{\bar k})_+(E_{11}W)&=   F(\bar\M_{\bar k},\bar\L_{\bar k})_+(E_{11}W)=E_{11}(F(\bar M,\bar L^{-1})\bar C_{kk})_-W,\\
           F(\M_{\bar k},\L_{\bar k})_+(E_{11}\bar W)&=  F(\bar\M_{\bar k},\bar\L_{\bar k})_+(E_{11}\bar W)=
           E_{11}(F(\bar M,\bar L^{-1})\bar C_{kk})_-\bar W.
      \end{aligned}
      \right.&
    \end{align}
\end{pro}
\begin{proof}
See Appendix B.
\end{proof}

If we set $F(x,y)=y^j$ in Proposition \ref{tec.pro}  and recall that
\begin{align*}
 \partial_{ja}W&=B_{ja}W, &  \partial_{ja}\bar W&=B_{ja}\bar W,
\end{align*}
with   $  B_{jk}=(C_{kk}L^j)_+$,  $B_{j\bar k}=(\bar C_{kk}\bar L^{-j})_-$  \cite{nuestro} we deduce
\begin{theorem} \label{lax whitham} The following scalar
linear systems hold
\begin{align}\label{evolutions}
    \partial_{ja}(E_{11}W)&=(\L^j_a)_+(E_{11}W)=(\bar\L^j_a)_+(E_{11}W),& \partial_{ja}(E_{11}\bar W)&=(\L^j_a)_+(E_{11}\bar W)=
    (\bar\L^j_a)_+(E_{11}\bar W)
  \end{align}
\end{theorem}

The linear system \eqref{evolutions} determines a set of commuting
flows for $(W,\bar W)$ which, as we will show in the next Section,
leads to the Whitham hierarchy in the dispersionless limit. For that
reason this system will be referred to as the dispersive Whitham
hierarchy of flows.

\subsection{Additional symmetries and string equations}
Using  Proposition \ref{pro:external parameters} we deduce the
following results on the additional symmetries
\begin{pro}\label{pro:stringy}
Given  an additional symmetry
\begin{align}\label{external dependence}
 \begin{aligned}
  \partial_{\tb}E_{11} W&=
  -E_{11}\bigg(\sum_{k=1}^N\big(F_k(M,L)C_{kk}- F_{\bar k}(\bar M,\bar L^{-1})
  \bar C_{kk}\big)\bigg)_-\cdot W,\\
  \partial_{\tb}E_{11}\bar W&=E_{11}\bigg(\sum_{k=1}^N\big(F_k(M,L)C_{kk}- F_{\bar k}(\bar M,\bar L^{-1})\bar C_{kk}\big)\bigg)_+\cdot\bar W,
\end{aligned}
\end{align}
then we have
\begin{align*}
  \partial_{\tb}(\Psi_a)&=-F_a(\M_a,\L_a)(\Psi_a)+\Big(\sum_{a'\in\mathcal S} F_{a'}(\M_{a'},\L_{a'})_+\Big)(\Psi_a)\\
  &=-F_a(\bar \M_a,\bar\L_a)(\Psi_a)+\Big(\sum_{a'\in\mathcal S} F_{a'}(\bar\M_{a'},\bar\L_{a'})_+\Big)(\Psi_a).
\end{align*}
\end{pro}
\begin{proof}
 From \eqref{external dependence} we get
\begin{align*}
  \partial_{\tb} E_{11}W&=-\sum_{k=1}^NE_{11}F_k(M,L)C_{kk}\cdot W+E_{11}\Big[\bigg(\sum_{k=1}^NF_k(M,L)C_{kk}\bigg)_+
  +\bigg(\sum_{k=1}^N\bar F_k(\bar M,\bar L^{-1})\bar C_{kk}\bigg)_-\Big]\cdot W,\\
   \partial_{\tb} E_{11}\bar W&=-\sum_{k=1}^NE_{11}\bar F_{\bar k}(\bar M,\bar L^{-1})\bar C_{kk} \cdot\bar   W +E_{11}\Big[\bigg(\sum_{k=1}^NF_k(M,L)C_{kk}\bigg)_+
  +\bigg(\sum_{k=1}^N F_{\bar k}(\bar M,\bar L^{-1})\bar C_{kk}\bigg)_-\Big]\cdot \bar W.
\end{align*}
Now, from Propositions \ref{pro: shift orlov} and \ref{tec.pro} we
conclude that
\begin{align*}
  \partial_{\tb}( E_{11}W)&=-\sum_{k=1}^NF_k(\M_k,\L_k)(W_{1k})E_{1k}+\bigg(\sum_{k=1}^N ( F_k(\M_k,\L_k)_++ F_{\bar k}(\M_{\bar k},\L_{\bar k})_+\bigg)(E_{11}W)\\
  &=-\sum_{k=1}^NF_k(\bar\M_k,\bar\L_k)(W_{1k})E_{1k}+\bigg(\sum_{k=1}^N ( F_k(\bar\M_k,\bar\L_k)_++ F_{\bar k}(\bar\M_{\bar k},\bar\L_{\bar k})_+\bigg)(E_{ll}W),\\
  \partial_{\tb}  (E_{11}\bar W)&=-\sum_{k=1}^N\bar F_k(\M_k,\L_k)(\bar   W_{1k})E_{lk} +\bigg(\sum_{k=1}^N ( F_k(\M_k,\L_k)_++ F_{\bar k}(\M_{\bar k},\L_{\bar k})_+\bigg) (E_{11}\bar W)\\
  &=-\sum_{k=1}^N\bar F_k(\bar\M_k,\bar\L_k)(\bar   W_{1k})E_{1k }
  +\bigg(\sum_{k=1}^N ( F_k(\bar\M_k,\bar\L_k)_+ +F_{\bar k}(\bar\M_{\bar k},\bar\L_{\bar k})_+\bigg) (E_{11}\bar W).
\end{align*}
and the result follows.
\end{proof}

As a consequence we have
\begin{pro}\label{pro:strings}
  If the string equation
  \begin{align}
  \label{string}
  E_{11}\sum_{k=1}^NF_k(M,L)C_{kk}=  E_{11}\sum_{k=1}^N F_{\bar k}(\bar M,\bar L^{-1})\bar C_{kk}
\end{align}
is satisfied, then
\begin{align*}
F_a(\M_a,\L_a)(\Psi_a)&=\Big(\sum_{a'\in\mathcal S}
F_{a'}(\M_{a'},\L_{a'})_+\Big)(\Psi_a),&
 F_a(\bar \M_a,\bar\L_a)(\Psi_a) &=\Big(\sum_{a'\in\mathcal S} F_{a'}(\bar\M_{a'},\bar\L_{a'})_+\Big)(\Psi_a),
\end{align*}
for all $a\in\mathcal S$.
\end{pro}
\begin{proof}
The string equations \eqref{string} imply the invariance conditions
\begin{align}
  \label{invariance}
  \partial_{\tb}E_{11}W=  \partial_{\tb}E_{11}\bar W=0.
\end{align}
Now, recalling Proposition \ref{pro:stringy} we get the desired
result.
\end{proof}

\section{The dispersionless limit}

We consider here the dispersionless limit of the multi-component 2D
Toda hierarchy. For that aim we use the vector wave functions
\eqref{PSI} at a given fixed value $n_0$ of the discrete variable $n$.
Thus, from Theorem \ref{lax whitham} the following auxiliary linear
system follows
  \begin{align}\label{evolutions baker}
    \partial_{ja}(\Psi_b)&=(\L_{a}^j)_+(\Psi_b)=(\bar \L_{a}^j)_+(\Psi_b)& a\in\mathcal S, j=1,2,\dots.
  \end{align}
Let us now introduce slow variables by
\begin{align*}
t_{\s,ja}&=\epsilon t_{ja},& s_{\s,a}&=\epsilon s_a,
\end{align*}
where $\epsilon$ is  a small real parameter and $s_{\s,a}$ are
assumed to be continuous variables. For the sake of simplicity,
we will henceforth denote by $(t_{ja},s_a)$ these slow variables.
Moreover, we assume that the wave functions have the
quasi-classical form
\begin{align*}
\Psi_a&=\exp\Big(\frac{\mathscr S_a}{\epsilon}\Big),& \mathscr
S_a&=\mathscr S_{a,0}+\epsilon\mathscr S_{a,1}+\cdots.
\end{align*}
with
\begin{align*}
  \mathscr{S}_a& =\mathcal{T}_{a}+\begin{cases}
   \epsilon\varphi_{1,11}z^{-1}+O(z^{-2}) &a=1,\\
   \epsilon\log\varphi_{1,k1}+O(z^{-1}) &a=k\neq 1\\
    \epsilon\log\bar\varphi_{0,k1}+O(z) &a=\bar k.\
  \end{cases}&
  \mathcal{T}_{a}&:=\begin{cases}
    (\epsilon n_{0} +s_{1})\log z+\sum_{j=1}^\infty  t_{ jl}z^j, & a=1,\\
  (\epsilon n_{0}+s_{ k}-\epsilon)\log z+\sum_{j=1}^\infty  t_{ jk}z^j, & a=k\neq 1,\\
(\epsilon n_{0}-  s_{ \bar k})\log z+\sum_{j=1}^\infty  t_{ j\bar
k}z^{-j}, & a=\bar k.
  \end{cases}
\end{align*}
 From these expressions we  deduce that as $\epsilon\to 0$
\begin{align*}
  \varphi_{1,11}&=O(\epsilon^{-1}), & \\
  \log\varphi_{1,1k}&=O(\epsilon^{-1}), &k\neq 1,\\
\log\bar\varphi_{0,1\bar k}&=O(\epsilon^{-1}). &
\end{align*}
As a consequence the coefficients in the operators $\L_a,\bar\L_a$
are Taylor series in $\epsilon$ while those of the Orlov--Schulman
operators $\M_a,\bar \M_a$ have at most a simple pole in
$\epsilon=0$.

We introduce some new variables
\begin{align*}
 &\begin{aligned}
 \sigma_{a}&:=s_{ a}, & a\neq 1,\\
  \sigma_{1}&:=\sum_{a\in\mathcal S}s_{ a},
 \end{aligned}
 &
&
\begin{aligned}
\bar \sigma_{a}&:=s_{ a}, & a\neq \bar 1,\\
  \bar\sigma_{1}&:=\sum_{a\in\mathcal S}s_{ a},
 \end{aligned}
\end{align*}
Observe that
\begin{align*}
  \frac{\partial}{\partial \sigma_{a}}&=\frac{\partial}{\partial s_{ a}}-\frac{\partial}{\partial s_{ 1}}, & a&\neq 1,&
  \frac{\partial}{\partial \bar\sigma_{a}}&=\frac{\partial}{\partial s_{ a}}-\frac{\partial}{\partial s_{ \bar 1}},
  & a&\neq \bar 1.
\end{align*}
The zero charge condition implies that $\sigma_1=\sigma_{\bar 1}=0$. Then, we define
\begin{align*}
\partial_{a}&:=\begin{cases}
\dfrac{\partial}{\partial  \sigma_{a}},& a\neq 1,\\[10pt]
-\dfrac{\partial}{\partial  \sigma_{a_0}},& a=1,
\end{cases},&
\bar\partial_{a}&:=\begin{cases}
\dfrac{\partial}{\partial  \bar\sigma_{a}},& a\neq \bar 1,\\[10pt]
-\dfrac{\partial}{\partial  \bar\sigma_{a_0}},& a=\bar 1.
\end{cases}
\end{align*}
Notice that
 \begin{pro}
   In the limit $\epsilon\to 0$ we have that
   \begin{align*}
  \Tp_{a}^j(\exp(\mathscr S_b/\epsilon))&=\exp(\Tp_{a}^j(\mathscr S_{b})/\epsilon)=
  \exp(j\partial_{a}(\mathscr S_{b,0})+O(\epsilon))\exp(\mathscr S_b/\epsilon),\\
   \bar \Tp_{a}^j(\exp(\mathscr S_b/\epsilon))&=\exp(\bar\Tp_{a}^j(\mathscr S_{b})/\epsilon)=
  \exp(j\bar\partial_{a}(\mathscr S_{b,0})+O(\epsilon))\exp(\mathscr S_b/\epsilon),\\
  \partial_{ja}(\exp(\mathscr S_b/\epsilon))&=(\partial_{ ja}(\mathscr S_{b,0})+O(\epsilon))\exp(\mathscr S_b/\epsilon).
\end{align*}
 \end{pro}

\subsection{Hamilton--Jacobi equations and dispersionless Whitham hierarchies}

As $\epsilon\to 0$ it follows that
\begin{align*}
 (\L_{a})^j_+(\Psi_b)&=\Big(\Pp_{ja}\big(\Exp{\partial_a\Ss_{b,0}}\big)+O(\epsilon)\Big)\Psi_b,\\
  (\bar{\L}_{a})^j_+(\Psi_b)&=\Big(\bar\Pp_{ja}\big(\Exp{\partial_a\Ss_{b,0}}\big)+O(\epsilon)\Big)\Psi_b,
\end{align*}
where $\Pp_{ja}$ and $\bar \Pp_{ja }$ are polynomials
 \begin{align*}
    \Pp_{j1}(Z)&=Z^{j}+P_{j1,j-1}Z^{j-1}+\dots+P_{j1,0},\\
     \Pp_{ja}(Z)&=P_{ja,j}Z^j+\dots+P_{ja,1}Z,& a\neq 1\\
         \bar \Pp_{j\bar 1}(Z)&=\bar P_{j\bar 1,j}Z^{j}+\bar P_{j\bar 1,j-1}Z^{j-1}+
    \dots+\bar P_{j\bar 1,1}Z-(1-\delta_{1a_0})\sum_{i=1}^j\bar P_{j\bar 1,i},\\
      \bar \Pp_{ja}(Z)&=\bar P_{ja,j}Z^j+\dots+\bar P_{ja,1}Z-(1-\delta_{a1})\sum_{i=1}^j\bar P_{ja,i}, & a\neq \bar 1.
  \end{align*}
Hence, as $\epsilon\to 0$
 we get from \eqref{evolutions baker} the  following
Hamilton--Jacobi type equations
\begin{pro}
  The following equations holds
  \begin{align}
    \partial_{ ja}(\Ss_{b,0})&=\Pp_{ja}\big(\Exp{\partial_a\Ss_{b,0}}\big),\label{HJ}\\
    \partial_{ ja}(\Ss_{b,0})&=\bar\Pp_{ja}\big(\Exp{\bar\partial_a\Ss_{b,0}}\big).\label{barHJ}
      \end{align}
\end{pro}
Next we show how these equations lead tot he two pictures of the Whitham
hierarchy described in the Appendix A.

\subsubsection{KP and Toda dispersionless limits from the Hamilton--Jacobi equations}

From the basic equation
\begin{align*}
  \frac{\partial\Psi_b}{\partial t_{11}}=(\L_1)_+(\Psi_b),
\end{align*}
 we get the important formula
\begin{align}\label{formula}
  \partial_{t_{ 11}}(\mathscr S_{b,0})&=
  \Exp{(\partial_{s_{ 1}}-\partial_{s_{ a}})(\Ss_{b,0})}+q_a,
  &
  &a\neq 1.
\end{align}
Where $q_a$ is an appropriate function defined in terms of derivatives of the leading coefficient of $\varphi_{1,11}$.
Observe that a family of equations as \eqref{formula} only occurs for the time $t_{11}$ and
 not for the times $t_{1a}$ with $a\neq 1$. This is a consequence of the fact that we have chosen
 the first row in the matrix wave functions, and we are dealing with the shifts of type $\Tp_a$.

\paragraph{The KP-picture  dispersionless limit }

\begin{definition}
 We introduce the dispersionless Lax functions in the KP picture, $z_a=z_a(\bs,\bt)$ by the implicit relations
\begin{align*}
p&=\partial_x\mathscr S_{a,0}(  z_a), &x:=t_{11},
\end{align*}
and the corresponding  dispersionless Orlov--Schulman functions by
\begin{align*}
m_a:= \frac{\partial \Ss_{a,0}}{\partial z}\bigg|_{z=z_a}.
\end{align*}
\end{definition}

This definition implies
\begin{align*}
  \Exp{\partial_1\Ss_{1,0}}\Big|_{z=z_1}&=p-q_{a_0},&\Exp{\partial_a\Ss_{a,0}}
  \big|_{z=z_a}&=\frac{1}{p-q_a},& a\neq 1.
\end{align*}

The next Proposition exhibits the asymptotic form of these functions
\begin{pro}\label{z-kp}
  The dispersionless Lax and Orlov--Schulman functions satisfy
  \begin{align*}
    z_a^{\sg a}&=\left\{
      \begin{aligned}
     & p+\ell_{1,0}+O(p^{-1}),&p&\to\infty,&a&=1,\\
      &\dfrac{\ell_{a,1}}{p-q_a}+\ell_{a,0}+O(p-q_a),&p&\to q_a & a&\neq 1,
      \end{aligned}
    \right.\\
    m_a&=
    (n_0+\sg(a)s_{ a})z_a^{-1}+
   \sum_{j=1}^\infty jt_{ j a}z_a^{j-1}+ z_a^{-1}
 \begin{cases}
     \sum_{j=1}^\infty \mu_{aj}(p-q_{a'})^{-j}, & a= 1,\\
      \sum_{j=1}^\infty \mu_{aj}(p-q_{a})^j, & a\neq 1.
    \end{cases}
  \end{align*}
\end{pro}
\begin{proof}
Particular cases of \eqref{bnota1} are
\begin{align*}
  \M_a(\psi_a)&=z\frac{\d\psi_a}{\d z},&   \L_a(\psi_a)&=z^{\sg a}\psi_a,
\end{align*}
which together with \eqref{ttlax} and \eqref{orlov whitham
expressions} imply
\begin{align*}
z^{\sg a}&=\begin{cases}
  \Exp{\partial_l\Ss_{1,0}}+\ell_{1,0}+\ell_{1,-1} \Exp{-\partial_l\Ss_{1,0}}+\ell_{1,-2} \Exp{-2\partial_l\Ss_{1,0}}+\cdots,& a=1\\
\ell_{a,1} \Exp{\partial_a\Ss_{a,0}}+\ell_{a,0}+\ell_{a,-1} \Exp{-\partial_a\Ss_{a,0}}+\ell_{a,-2} \Exp{-2\partial_a\Ss_{a,0}}+\cdots,& a\neq 1,
\end{cases}\\
 \frac{\partial \Ss_{a,0}}{\partial z}&=(n_0+\sg(a)s_{ a})z^{-1}+
   \sum_{j=1}^\infty jt_{ j a}z^{j-1}+ z^{-1}\sum_{j=1}^\infty \mu_{aj}\Exp{-j\partial_a \Ss_{a,0}}
\end{align*}
and the evaluation at $z=z_a$ gives the desired result.
\end{proof}

Therefore for $a\neq 1$ we have
\begin{align*}
(  \partial_a(\mathscr S_{b,0}))\Big|_{z=z_b}&=-\log(p-q_{a}), & a\neq 1\\
  (\partial_{ ja}(\mathscr S_{b,0}))\Big|_{z=z_b}&=\Pp_{ja}\Big(\frac{1}{p-q_{a}}\Big)=:\Omega_{ja},& a\neq 1\\
  (\partial_{ j1}(\mathscr S_{b,0}))\Big|_{z=z_b}&=\Pp_{j1}(p-q_{a'})=:\Omega_{j1}, &j>1
  \end{align*}
Then we have that
\begin{align*}
  \d \Ss_{b,0}=m_b\d z_b+p\,\d x-\sum_{a\neq 1}\log(p-q_{a})\d s_{ a}+{\sum_{j,a}}^\prime\Omega_{ja}\d t_{ ja}
\end{align*}
where the $\Sigma'$ indicates the sum over the set of indexes $(j,a)$
where $j=1,2,\cdots$ and $a\in\mathcal S$ excluding the case $j=1$
and $a=l$. Thus the functions $\d \Ss_{b,0}$ determine a solution of
the zero-genus Whitham hierarchy with $2N$ punctures in the KP
picture (see Appendix A).

\paragraph{The Toda-picture dispersionless limit}
We again consider equation \eqref{formula}
\begin{align*}
 \partial_{t_{ 11}}(\mathscr S_{b,0})&=
  \Exp{(\partial_{s_{ 1}}-\partial_{s_{ a}})(\Ss_{b,0})}+q_a,
\end{align*}
which implies
\begin{align*}
 \Exp{-\partial_{a_0}\mathscr S_{b,0}}&= \Exp{-\partial_{a}\mathscr S_{b,0}}+Q_{a},& a,a_0&\neq 1, & Q_a&:=q_a-q_{a_0}
\end{align*}

\begin{definition}
In the Toda representation the dispersionless Lax function $z_a=z_a(\bs,\bt)$ is given by the implicit relation
\begin{align*}
p&=\Exp{-\partial_{x}\mathscr S_{b,0}}\Big|_{z=  z_b},& x:=-\sigma_{a_0},
\end{align*}
and the dispersionless Orlov--Schulman function by
\begin{align*}
m_a:= z\frac{\partial \Ss_{a,0}}{\partial z}\bigg|_{z=z_a}.
\end{align*}
\end{definition}

Observing that
\begin{align*}
  \Exp{\partial_a\Ss_{b,0}}&=\frac{1}{\Exp{-\partial_{a_0}\Ss_{b,0}}-Q_a}
\end{align*}
we conclude
\begin{align}\label{exp-toda}
  \Exp{\partial_x\Ss_{1,0}}\Big|_{z=z_1}&=p,&
\Exp{\partial_{a_0}\Ss_{a_0,0}}\Big|_{z=z_{a'}}&=p^{-1},&
\Exp{\partial_a\Ss_{a,0}}\Big|_{z=z_a}&=\frac{1}{p-Q_a},& a_0&\neq 1& a&\neq 1,a_0.
\end{align}

Hence, we deduce
\begin{pro}
  The dispersionless Lax and Orlov--Schulman functions  in the Toda-picture dispersionless limit satisfy
  \begin{align*}
    z_a^{\sg a}&=\left\{\begin{aligned}
     &p+\ell_{1,0}+O(p^{-1}),& p&\to \infty,&a&=1,\\
     &\ell_{2,1} p^{-1}+O(1),& p&\to 0, &a&=a_0,\\
      &\dfrac{\ell_{a,1}}{p-Q_a}+O(1),& p&\to Q_a, & a&\neq 1,a_0.
    \end{aligned}
          \right.\\
    m_a&=
    (n_0+\sg(a)s_{ a})+
   \sum_{j=1}^\infty jt_{ j a}z_a^{j}+
 \begin{cases}
     \sum_{j=1}^\infty \mu_{1j}p^{-j}, & a= 1,\\
     \sum_{j=1}^\infty \mu_{aj}p^{j}, & a= a_0,\\
   \sum_{j=1}^\infty \mu_{aj}(p-Q_{a})^j, & a\neq 1,a_0.
    \end{cases}
  \end{align*}
\end{pro}
\begin{proof}
Proceed as in the proof of Proposition \ref{z-kp} and use \eqref{exp-toda}.
\end{proof}

As in the KP case we get now
\begin{align*}
(  \partial_a(\mathscr S_{b,0}))\Big|_{z=z_b}&=-\log(p-Q_{a}), & a\neq 1,a_0,\\
  (\partial_{ ja}(\mathscr S_{b,0}))\Big|_{z=z_b}&=\Pp_{ja}\Big(\frac{1}{p-Q_{a}}\Big)=:\Omega_{ja},& a\neq 1,a_0,\\
  (\partial_{ j1}(\mathscr S_{b,0}))\Big|_{z=z_b}&=\Pp_{j1}(p)=:\Omega_{j1}, &\\
  (\partial_{ ja_0}(\mathscr S_{b,0}))\Big|_{z=z_b}&=\Pp_{ja_0}\Big(p^{-1}\Big)=:\Omega_{ja_0}.
  \end{align*}

Hence we have that
\begin{align*}
  \d \Ss_{b,0}=m_b\d \log z_b+\log p\,\d x-\sum_{a\neq 1,a'}\log(p-Q_{a})\d s_{ a}+{\sum_{j\geq 1,a\in\mathcal S}}\Omega_{ja}\d t_{ ja},
\end{align*}
and therefore the functions $\Ss_{b,0}$ determine a solution of the
zero-genus Whitham hierarchy with $2N$ punctures in the Toda picture
(see Appendix A).

\paragraph{An alternative Toda-picture dispersionless limit}
 From the basic equation
\begin{align*}
  \frac{\partial\Psi_b}{\partial t_{1\bar 1}}=(\bar\L_{\bar 1})_+(\Psi_b),
\end{align*}
we deduce
\begin{align*}
  \partial_{ 1\bar 1}\Ss_{b,0}&= r_{1}\Exp{(\partial_{ s_{\bar 1}}-\partial_{ s_{ 1}})(\Ss_{b,0})}=
  r_{a}\big(\Exp{(\partial_{ s_{\bar 1}}-\partial_{ s_{ a}})(\Ss_{b,0})}-1\big),&
  a\neq 1,\bar 1
\end{align*}
for some functions $r_a$. Hence,
\begin{align*}
  \Exp{\bar \partial_a\Ss_{b,0}}&=
  \frac{\Exp{\bar\partial_1\Ss_{b,0}}}{\Exp{\bar\partial_1\Ss_{b,0}}+\rho_a},&
  \rho_a&:=\frac{r_a}{r_1}.
\end{align*}

Now, we take $a_0=1$  and define
\begin{definition}
  The  dispersionless Lax functions $ z_a$ are defined by the
  implicit relation
\begin{align*}
  \partial_x\Ss_{a,0}|_{z= z_a}&=\log p, & x&=-\bar\sigma_1
\end{align*}
while the dispersionless Orlov--Schulman functions are defined by
\begin{align*}
 m_a:= z_a\frac{\partial \Ss_{a,0}}{\partial
z}\bigg|_{z= z_a}.
\end{align*}
\end{definition}
Observe that
\begin{align*}
   \Exp{\partial_1\Ss_{1,0}}\Big|_{z= z_1}&=p,&
      \Exp{\partial_{\bar 1}\Ss_{\bar 1,0}}\Big|_{z= z_{\bar 1}}&=p^{-1},&
    \Exp{\partial_a\Ss_{a,0}}\Big|_{z= z_a}&=  \frac{p}{p+\rho_a}, &a\neq 1,\bar 1.
\end{align*}
\begin{pro}
  The dispersionless Lax and Orlov--Schulman functions are of the following form
  \begin{align*}
     z_a^{\sg a}&=\left\{
     \begin{aligned}
      &p+\bar \ell_{1,0}+O(p^{-1}),& p&\to \infty,&a&=1\\
        &\bar  \ell_{\bar 1,1}p^{-1}+\bar \ell_{\bar 1,0}+O(p),&p&\to 0,&a&=\bar 1\\
      &\bar \ell_{a,1}  \dfrac{p}{p+\rho_a}+\bar \ell_{a,0}+O\Big(\Big(  \dfrac{p}{p+\rho_a}\Big)^{-1}\Big), & p&\to -\rho_a,&a&\neq 1,
     \end{aligned}
          \right.\\
    m_a&=
    (n_{0}+\sg(a)s_{ a})+
   \sum_{j=1}^\infty jt_{ j a}\bar z_a^{j-1}+
 \begin{cases}
     \sum_{j=1}^\infty \mu_{aj}p^{-j}, & a= 1,\\
     \sum_{j=1}^\infty \mu_{a j}p^{j}, & a= \bar 1,\\
     \sum_{j=1}^\infty \mu_{aj}\Big(  \dfrac{p}{p+\rho_a}\Big)^j, & a\neq 1,\bar 1.
    \end{cases}
  \end{align*}
\end{pro}
\begin{proof}
Particular cases of \eqref{bnota1} are
\begin{align*}
  \bar \M_a(\Psi_a)&=z\frac{\d\Psi_a}{\d z},&  \bar  \L_a(\Psi_a)&=z^{\sg a}\Psi_a,
\end{align*}
which together with \eqref{ttlax} and \eqref{orlov whitham
expressions} imply
\begin{align*}
z^{\sg a}&=\begin{cases}
  \Exp{\bar\partial_1\Ss_{1,0}}+\bar \ell_{1,0}+\bar \ell_{1,-1}
  \Exp{-\bar\partial_1\Ss_{1,0}}+\bar \ell_{1,-2} \Exp{-2\bar\partial_1\Ss_{1,0}}+\cdots,& a=1,\\
\bar\ell_{a,1} \Exp{\bar\partial_a\Ss_{a,0}}+\bar\ell_{a,0}+\bar
\ell_{a,-1} \Exp{-\bar\partial_a\Ss_{a,0}}+\bar\ell_{a,-2}
\Exp{-2\bar\partial_a\Ss_{a,0}}+\cdots,& a\neq 1,
\end{cases}\\
z \frac{\partial \Ss_{a,0}}{\partial z}&=(n_{0}+\sg(a)s_{ a})+
   \sum_{j=1}^\infty jt_{ j a}z^{j}+ \sum_{j=1}^\infty\bar  \mu_{aj}\Exp{-j\bar\partial_a
   \Ss_{a,0}}.
\end{align*}
and the evaluation of these expressions at $z= z_a$ gives the
result.
\end{proof}

Therefore,
\begin{align*}
  \partial_x(\Ss_{b,0})\Big|_{z= z_b}&=\log p,\\
\partial_a(\Ss_{b,0})\Big|_{z= z_b}&= \log\Big(  \frac{p}{p+\rho_a}\Big), &a\neq 1,\bar 1,\\
  (\partial_{ ja}(\mathscr S_{b,0}))\Big|_{z=
  z_b}&=\bar\Pp_{ja}
  \Big(  \frac{p}{p+\rho_a}\Big)=: \Omega_{ja},& a\neq 1,\bar 1,\\
    \partial_{ j\bar 1}(\Ss_{b,0})\Big|_{z= z_b}&=\bar\Pp_{j\bar 1}(p^{-1})
    =:\Omega_{j\bar 1},&
    \\
  (\partial_{ j1}(\mathscr S_{b,0}))\Big|_{z= z_b}&=\bar\Pp_{j1}(p)=:\Omega_{j1},
  &j>1.
  \end{align*}
In this way we have
\begin{align*}
  \d \Ss_{b,0}= m_b\d \log z_b+\log(p)\d x+\sum_{a\neq \bar 1,1}
  \log\Big( \frac{p}{p+\rho_a}\Big)\d s_{ a}+\sum_{j\geq 1,a\in\mathcal S}\Omega_{ja}\d
  t_{ ja}.
\end{align*}
As we will show at the end of this section the functions $\Ss_{b,0}$
determine a solution of the zero-genus Whitham hierarchy with $2N$
punctures in the Toda picture.

\subsection{The dispersionless limits of the string equations}

Let us consider operators of the form
\begin{align*}
  F_a(\M_a,\L_a)&=\sum_{i\geq 0,j\in\Z}F_{aij}\M_a^i\L_a^j,&
   F_a(\bar \M_a,\bar\L_a)=\sum_{i\geq 0,j\in\Z}F_{aij}\bar
  \M_a^i\bar\L_a^j.
  \end{align*}
In order to formulate their  dispersionless limits it is convenient
to assume that the coefficients satisfy as $\epsilon\to 0$ that
\begin{align*}
  F_{aij}=F_{aij,0}\epsilon^i+O(\epsilon^{i+1}).
\end{align*}
Recalling \eqref{bnota1} and observing that
\begin{align*}
\Big(z\frac{\d}{\d z}\Big)^i&=z^{i}\frac{\d^i}{\d z^i}+\sum_{i'=2}^i\binom{i'}{2}z^{i'-1}\frac{\d^{i'-1}}{\d z^{i'-1}},\\
\frac{\partial^i\Psi_a}{\partial
z^i}&=\Big(\epsilon^{-i}\Big(\frac{\partial\Ss_{a,0}}{\partial
z}\Big)^i+O(\epsilon^{-i+1})\Big)\Psi_a
\end{align*}
we get
\begin{align*}
  (\Psi_a)\overleftarrow{F_a\Big(z\frac{\d}{\d z},z^{\sg a}\Big)}=
  \Bigg[\sum_{i\geq 0, j\in\Z}F_{aij,0}z^{\sg ( a)j}\Big(\frac{\partial \Ss_{a,0}}{\partial z}\Big)^i+O(\epsilon^{})\Bigg]\Psi_a.
\end{align*}
Hence,
\begin{align*}
F_{a,0}(z_a,m_a):= \bigg[\lim_{\epsilon\to 0}  (\Psi_a)\overleftarrow{F_a\Big(z\frac{\d}{\d z},z^{\sg a}\Big)}\Psi_a^{-1}\bigg]_{z=z_a}
 &=
\left\{
\begin{aligned}
& \sum_{i\geq 0, j\in\Z}F_{aij,0}z_a^{i+\sg ( a)j}m_a^i,&\text{KP},\\
 &\sum_{i\geq 0,
 j\in\Z}F_{aij,0} z_a^{\sg ( a)j} m_a^i,& \text{Toda.}
\end{aligned}
\right.
\end{align*}
We define
\begin{align*}
 F_{a,0+}&:=\begin{cases}
  \Big[\lim_{\epsilon\to 0}
  \dfrac{F_a(\M_a,\L_a)_+(\Psi_a)}{\Psi_a}
  \Big]_{z=z_a},&\text{unbared cases}\\[10pt]
  \Big[\lim_{\epsilon\to 0}
  \dfrac{F_a(\bar\M_a,\bar\L_a)_+(\Psi_a)}{\Psi_a}
  \Big]_{z= z_a},&\text{bared cases}
 \end{cases}
\end{align*}

\begin{pro}
Given
\begin{align*}
  &F_a(\M_a,\L_a)=\sum_{j\in\Z}f_{aj}\Tp_a^j,&&\text{ or }  & F_a(\bar\M_a,\bar\L_a)=\sum_{j\in\Z}\bar f_{aj}\bar\Tp_a^j,
\end{align*}
with $f_{ai}=f_{ai|0}+O(\epsilon)$ as $\epsilon\to 0$,
their dispersionless limits are
\begin{align*}
  F_{a,0}&=
  \left\{
  \begin{aligned}
    &\begin{cases}
    \sum_{j\in\Z}\dfrac{f_{aj|0}}{(p-q_a)^j}, & a\neq 1,\\
    \sum_{j\in\Z} f_{1j|0}(p-q_{a'})^j & a=1,\end{cases}   &&\text{KP,}\\
    &\begin{cases}
     \sum_{j\in\Z}\dfrac{f_{aj|0}}{(p-Q_a)^j}, &a\neq 1,a',\\
       \sum_{j\in\Z}f_{1j|0}p^j, &a=1,\\
        \sum_{j\in\Z}f_{a'j|0}p^{-j}, &a=a',\\
        \end{cases}&&\text{Toda ,}\\
     & \begin{cases}
       \sum_{j\in\Z}f_{aj|0}\Big( \dfrac{p}{p+\rho_a}\Big)^j, & a\neq 1,\bar 1,\\\sum_{j\in\Z}f_{1j|0}p^j,& a=1,\\
      \sum_{j\in\Z}f_{\bar 1j|0}p^{-j},& a=\bar 1,
 \end{cases}&&\text{Alternative Toda.}
  \end{aligned}
  \right.
  \end{align*}
Moreover,
\begin{align*}
  F_{a,0+}&=
  \left\{
  \begin{aligned}
    &\begin{cases}
    \sum_{j>0}\dfrac{f_{aj|0}}{(p-q_a)^j}, & a\neq 1,\\
    \sum_{j\geq 0} f_{1j|0}(p-q_{a'})^j & a=1,\end{cases}   &&\text{ KP,}\\
    &\begin{cases}
     \sum_{j>0}\dfrac{f_{aj|0}}{(p-Q_a)^j}, &a\neq 1,a',\\
       \sum_{j\geq 0}f_{1j|0}p^j, &a=1,\\
        \sum_{j>0}f_{a'j|0}p^{-j}, &a=a',\\
        \end{cases}&&\text{Toda,}\\
     & \begin{cases}
       \sum_{j>0}f_{aj|0}\Big(\Big( \dfrac{p}{p+\rho_a}\Big)^j-1\Big), & a\neq 1,\bar 1,\\\sum_{j\geq 0}f_{1j|0}p^j,& a=1,\\
      \sum_{j>0}f_{\bar 1j|0}p^{-j},& a=\bar 1,
 \end{cases}&&\text{Alternative Toda.}
  \end{aligned}
  \right.
  \end{align*}
In particular
\begin{align*}
  \Omega_{ja}&=
  \begin{cases}
    (z_a^{\sg(a) j})_+,&\text{KP, Toda,}\\
(\bar z_a^{\sg(a)j})_+, &\text{Alternative Toda.}
  \end{cases}
\end{align*}
\end{pro}
\begin{proof}
The formulae follow from the identity
\begin{align*}
  (\Psi_a)\overleftarrow{F_a\Big(z\frac{\d}{\d z},z^{\sg a}\Big)}&= F_a(\M_a,\L_a)(\Psi_a)=\sum_{j\in\Z}f_{aj}\Exp{j\partial_a\Ss_{a,0}+O(\epsilon)}\Psi_a\\&=
   F_a(\bar \M_a,\bar\L_a)(\Psi_a)=\sum_{j\in\Z}f_{aj}\Exp{j\bar\partial_a\Ss_{a,0}+O(\epsilon)}\Psi_a.
\end{align*}
\end{proof}

As a consequence
\begin{pro}
 If the string equations \eqref{string} hold, their corresponding dispersionless limits
\begin{align}
  \label{dstring}
  F_{a,0}(z_a,m_a)&=\sum_{b\in\mathcal S}F_{b,0+},
   & \forall a\in\mathcal S
\end{align}
 are satisfied.
\end{pro}
\begin{proof}
It follows from Proposition \ref{pro:strings}.
\end{proof}

These dispersionless string equations are of the type considered in
\cite{string} for the dispersionless Whitham hierarchy.  Moreover, given
a decomposition  $\mathcal S=\mathcal I\cup \mathcal J$ into two
disjoint subsets, we may take
\begin{align*}
  P_{a,0}&=\begin{cases}
    z_a^{\ell_a}, & a\in \mathcal I,\\
   - \dfrac{m_a}{\ell_a z^{\ell_a-1}}, & a\in \mathcal J,
  \end{cases}& Q_{a,0}&=\begin{cases}
    \dfrac{m_a}{\ell_a z^{\ell_a-1}}, & a\in \mathcal I, \\   z_a^{\ell_a}, & a\in \mathcal J.  \end{cases}
\end{align*}
The corresponding dispersive string equations are
\begin{align*}
E_{11}(  \sum_{k\in (\mathcal I\cap \bS)}L^{\ell_k}C_{kk}-\sum_{k\in (\mathcal J\cap \bS)}
\ell_k^{-1}ML^{-\ell_k+1}C_{kk})&=
    E_{11}(\sum_{k\in (\mathcal I\cap \bar\bS)}\bar L^{-\ell_k}\bar C_{kk}-\sum_{k\in
     (\mathcal J\cap \bar \bS)}\ell_k^{-1}\bar M\bar L^{\ell_k-1}\bar C_{kk}),\\
 E_{11}     (\sum_{k\in (\mathcal I\cap \bS)}\ell_k^{-1}ML^{-\ell_k+1}C_{kk}+\sum_{k\in (\mathcal J\cap \bS)}
 L^{\ell_k}C_{kk})&=E_{11}(\sum_{k\in (\mathcal I\cap \bar \bS)}\ell_k^{-1}\bar M\bar L^{\ell_k-1}\bar C_{kk}+
    \sum_{k\in (\mathcal J\cap \bar\bS)}\bar L^{-\ell_k}\bar C_{kk}).
\end{align*}
If $\mathcal J=\varnothing$ we get
\begin{align*}
E_{11}  \sum_{k=1}^NL^{\ell_k}C_{kk}&=
    E_{11}\sum_{k=1}^N\bar L^{-\ell_k}\bar C_{kk},\\
      E_{11}\sum_{k=1}^N\ell_k^{-1}ML^{-\ell_k+1}C_{kk}&=E_{11}\sum_{k=1}^N\ell_k^{-1}\bar M\bar L^{\ell_k-1}\bar C_{kk}.
\end{align*}
For positive integers $\ell_a$ the dispersionless limits of these dispersive string equations describe the
algebraic orbits of the genus 0 Whitham hierarchy \cite{krichever}. The first of these  dispersive string
equations gives the multigraded reduction as discussed in \cite{nuestro}.

We return to the equivalence of the alternative Toda  and Toda
pictures. First we notice that within alternative Toda picture we have
Laurent expansions in $Z_a$
\begin{align*}
  Z_a&:=\dfrac{p}{p+\rho_a}=1-\rho_a \zeta_a,& \zeta_a&:=\frac{1}{p+\rho_a}.
\end{align*}
The functions $Z_a^j$ are singular at $p=-\rho_a$ and
$\lim_{p\to\infty}Z_a^j=1$, thus the linear combinations of factors
$Z_a^j-1$ which appear  in the construction
 of the $\Omega_{ja}$, lead to singular functions in $p=-\rho_a$ normalized to 0 at infinity.
Hence, if we express $z_a$ as Laurent series in $\zeta_a$, the function $\Omega_{ja}$
  is just the singular part corresponding to the projection to power series in $\zeta_a$ with
  non constant term. Thus, we recover the Toda picture of the genus 0  Whitham hierarchy (see Appendix A).

Given operators $P_a$ and $Q_a$ as in Proposition \ref{Lax C baker}
we get for the commutator
\begin{multline*}
  (\Psi_a)\overleftarrow{[P_a,Q_a]}\Big(z\frac{\d}{\d z},z^{\sg a}\Big)
  =(\Psi_a)\bigg[\overleftarrow{P_a\Big(z\frac{\d}{\d z},z^{\sg a}\Big)},
  \overleftarrow{Q_a\Big(z\frac{\d}{\d z},z^{\sg a}\Big)}\bigg]\\[6pt]=\bigg[\sum_{\substack{i_1,i_2\geq 0\\j_1,j_2\in\Z}}
  P_{ai_1j_1,0}Q_{ai_2j_2,0}\sg (a)(i_2j_1-i_1j_2)z^{\sg (a)(j_1+j_2)+i_1+i_2-1}\Big(\frac{\partial \Ss_{a,0}}{\partial}\Big)^{i_1+i_2-1}\epsilon+O(\epsilon^2)\bigg]\Psi_a
\end{multline*}
so that for the KP picture  we find
\begin{align*}
\bigg[\lim_{\epsilon\to 0}
(\Psi_a)\overleftarrow{\epsilon^{-1}[P_a,Q_a]}\Big(z\frac{\d}{\d
z},z\Big) \Psi_a^{-1}\bigg]_{z=z_a}&=\sum_{\substack{i_1,i_2\geq 0\\j_1,j_2\in\Z}}
  P_{ai_1j_1,0}Q_{ai_2j_2,0}\sg(a)(i_2j_1-i_1j_2)z_a^{\sg(a)(j_1+j_2)+i_1+i_2-1}m_a^{i_1+i_2-1}
  \\[6pt]&=    \{P_{0,a},Q_{0,a}\}_0,
 \end{align*}
while for the  Toda picture we have
\begin{align*}
\bigg[\lim_{\epsilon\to 0}
(\Psi_a)\overleftarrow{\epsilon^{-1}[P_a,Q_a]} \Big(z\frac{\d}{\d
z},z\Big)\Psi_a^{-1}\bigg]_{z= z_a}&=\sum_{\substack{i_1,i_2\geq 0\\j_1,j_2\in\Z}}
  P_{ai_1j_1,0}Q_{ai_1j_1,0}\sg(a)(i_2j_1-i_1j_2) z_a^{\sg(a)(j_1+j_2)} m_a^{i_1+i_2-1}
  \\[6pt]&=    \{ P_{0,a}, Q_{0,a}\}_1.
 \end{align*}
Thus,
\begin{align*}
  [P_a,Q_a]&=\epsilon& &\Longrightarrow &\{P_{a,0},Q_{a,0}\}_0&=1 \text{ or } \{ P_{a,0}, Q_{a,0}\}_1=1.
\end{align*}

\section*{Appendix A: Whitham hierarchies in the zero genus case}

The zero-genus Whitham hierarchies \cite{krichever} are systems of flows on a \emph{phase space} $\widehat{M}_0$ of data associated to algebraic Riemann surfaces of genus $0$. The points of  $\widehat{M}_0$ are given $(\Gamma,\,Q_a,\,z_a^{-1})$, where $\Gamma$ is an algebraic Riemann surface of genus $0$, $Q_a$ are   $N$ points (punctures) of $\Gamma$ and $z_a^{-1}$ denote local coordinates around each $Q_a$ such that $z_a^{-1}(Q_a)=0$. In order to formulate Whitham flows on  $\widehat{M}_0$ it is convenient to introduce  a meromorphic function $p=p(Q)$ on $\Gamma$ such that the local coordinates have asymptotic expansions of the form
\begin{gather}\label{1.1}\everymath{\displaystyle}
      z_a=\begin{cases}
        p+\ell_{1,0}+\sum_{n=1}^\infty \frac{\ell_{1,-n}}{p^n}, & a=1,\\\\
        \dfrac{\ell_{a,1}}{p-q_a}+\sum_{n=0}^\infty \ell_{a,-n} (p-q_a)^n,&
        a=2,\dots, N,
      \end{cases}
\end{gather}
where $p(Q_a)=q_a$ with $q_1=\infty$. In general the points of the
phase space $\widehat{M}_0$ are characterized by an infinite number
of parameters $\bw:=(w_i)$ of the set $(q_a, \ell_{a,n})$. However,
under appropriate reduction conditions on the form of $\Gamma$ only
a finite number  of these parameters are independent and constitute a
coordinate system for  $\widehat{M}_0$. \vspace{0.2cm} \noindent
\subsection*{Example: Algebraic orbits}

If we restrict to zero-genus Riemann surfaces $\Gamma$ of the form
\begin{equation}\label{ao}
\lambda=p^{n_1}+\sum_{n=0}^{n_1-1}\,u_{1n}\,p^n+\sum_{i=2}^N\,
\sum_{n=1}^{n_q}\dfrac{u_{in}}{(p-q_i)^n},
\end{equation}
we may take $Q_a=(\lambda_a,p_a)=(\infty,q_a)\,( a=1,\ldots,q),$ with corresponding local coordinates given by
\[
z_a=\lambda^{1/n_a}.
\]
The function $p(\lambda,p)=p$ is meromorphic on $\Gamma$ and the local coordinates have asymptotic expansions of the form \eqref{1.1}.  In this case the points of the phase space $\widehat{M}_0$ are characterized by the parameters $\bw=(\{q_a\}_{a=2}^N,\{u_{1n}\}_{n=0}^{n_1},\ldots,\,\{u_{Nn}\}_{N=1}^{n_N})$

\vspace{0.3cm}

The Whitham flows $\bw(\bt)=(w_i(\bt))$ are introduced through sets $\mathbf{\Omega}:=\{\Omega_A(\bw,\,p)\}$
 of functions, with meromorphic differentials $\partial_p\,\Omega_A(\bw,\,p)\,\d\,p$, which  satisfy the  conditions:
\begin{enumerate}
\item One of the  functions $\Omega_{A_0}$ is independent of the data $\bw$.
\item There exist local functions $S_a(\bt,z_a)$ around the
    punctures satisfying
\begin{equation}\label{wc}
\partial_A\,S_a(\bt,z_a)=\Omega_A(\bw(\bt),\,p(\bt,z_a)).
\end{equation}
Here $
\partial_A:=\partial / \partial t_A$ and $\bt$ denotes the set of flow parameters $t_A$.
\end{enumerate}
\noindent The first condition only demands to include  a function of the
form $\Omega(p)$ in $\mathbf{\Omega}$. On the other hand, it is
obvious that the second condition is satisfied if and only if the following
Zakharov--Shabat equations are satisfied
\begin{equation}\label{zs}
\partial_B\,\Omega_A-\partial_A\,\Omega_B+\{\Omega_A,\Omega_B\}=0,
\end{equation}
where $\partial_t:=\sum_i  \partial_t\,w_i\,\partial_{w_i}$ for $t=t_A,\,t_B$, and $\{\,,\,\}$ denotes the Poisson bracket
\[
\{F,G\}:=\omega(p)\,\Big(
\partial_p\,F\,\partial_x\,G-
\partial_x\,F\,\partial_p\,G
\Big),\quad \omega(p):=(\partial_p\,\Omega_{A_0}(p))^{-1},\quad x:=t_{A_0}.
\]
We may write \eqref{wc} as
\begin{equation}\label{wc1}
\d\, S_a=m_a\,\d \,\Omega_{A_0}(z_a)+\sum_A\,\Omega_A\,\d\,t_A,
\quad m_a:=\omega(z_a)^{-1}\,\dfrac{\partial S_a}{\partial z_a},
\end{equation}
which implies
\begin{equation}\label{wc2}
\d \,\Omega_{A_0}(z_a)\wedge \d\,m_a=
\sum_A\,\d\,\Omega_A\wedge\d\,t_A,
\end{equation}
and by equating the coefficients of $\d\, p\wedge\d\,x$ in both
members of \eqref{wc2} yields
\begin{equation}\label{pbr}
\{z_a,m_a\}=\omega(z_a).
\end{equation}
Moreover, if we identify the coefficients of $\d\, p\wedge\d\,t_A$ and $\d\, x\wedge\d\,t_A$ in \eqref{wc2} we get
\[
\begin{cases}
\partial_p\,z_a\,\partial_A\,m_a-
\partial_A\,z_a\,\partial_p\,m_a=\omega(z_a)\,\partial_p\,\Omega_A,\\
\partial_x\,z_a\,\partial_A\,m_a-
\partial_A\,z_a\,\partial_x\,m_a=\omega(z_a)\,\partial_x\,\Omega_A,
\end{cases}
\]
so that taking \eqref{pbr} into account we deduce the system of Lax equations
\begin{equation}\label{wc3}
\partial_A\,z_a=\{\Omega_A,z_a\},\quad \partial_A\,m_a=\{\Omega_A,m_a\}.
\end{equation}

\vspace{0.3cm}

As it was shown in \cite{mano}-\cite{mano1} important classes of solutions of the zero-genus Whitham
hierarchy  can be obtained from systems of canonical pairs of constrains  (string equations) of the form
\begin{equation}\label{2.1}
\begin{cases}
P_1(z_1,m_1)=P_2(z_2,m_2)=\cdots= P_{N}
(z_{N},m_{N}),\\
Q_1(z_1,m_1)=Q_2(z_2,m_2)=\cdots= Q_{N}
(z_{N},m_{N}),
\end{cases}
\end{equation}
where $(P_a,Q_a)$ are $N$ pairs of canonically conjugate functions
\begin{equation}\label{2.2}
\{P_a(p,x),Q_a(p,q)\}=\omega(p).
\end{equation}
In particular this type of methods applies for finding solutions for algebraic orbits. Indeed these solutions are associated to string equations generated by
\begin{equation}\label{ae1}
P_a(p,x)=p^{n_a},\quad Q_a(p,x)=\dfrac{x}{n_a\,p^{n_a-1}}+f_a(p).
\end{equation}

\subsection*{The KP picture}

The KP picture of the zero-genus Whitham hierarchy with $N$
punctures \cite{krichever} is formulated by assuming $\ell_{1,0}\equiv
0$ in the asymptotic expansions
 \eqref{1.1} and by taking the following functions $\Omega_A$
\begin{gather}\label{kp1}\everymath{\displaystyle}
\Omega_{na}:=\begin{cases}   (z_a^n)_{(a,+)} ,& n\geq 1
,\\\\
-\log(p-q_a), &n=0,\quad a=2,\dots,N.
\end{cases}
\end{gather}
Here $(\cdot)_{(a,+)}$ stand for the projectors on the subspaces
generated by $\{p^n\}_{n=0}^\infty$ (case $a=1$) and
$\{(p-q_a)^{-n}\}_{n=1}^\infty$ (cases $a\geq 2$). In this case
\[
A_0=(1,1),\quad x=t_{1,1},\quad \Omega_{A_0}=p,
\]
and the Poisson bracket is given by
\[
\{F,G\}:=
\partial_p\,F\,\partial_x\,G-
\partial_x\,F\,\partial_p\,G
.
\]
The functions $\Omega_A$ satisfy the compatibility conditions \eqref{zs} so that there exist functions $S_a$ such that
\begin{equation}\label{wckp1}
\d\, S_a=m_a\,\d \,z_a+p\,\d\,x-\sum_{a\neq 1}\,\log(p-q_a)\,\d\,t_{0a}+\sum_{a=1}^N\sum_{n\geq 1}\,\Omega_{na}\,\d\,t_{na}
.
\end{equation}

\subsection*{The Toda picture}

A simple redefinition of the meromorphic function $p(Q)$ used to
define the KP flows of the Whitham hierarchy with $N$ punctures
supplies a different picture (the Toda picture) of the hierarchy. Indeed if
we set
\[
p_{\text{Toda}}=p_{\text{KP}}-q_{a_0},
\]
for a given index $a_0$, then now $u_{1,1}=q_{a_0}$ and we may take
\[
A_0:=(0,a_0),\quad x:=-t_{0,a_0},\quad \Omega_{A_0}:=-\log\,p.
\]
Thus the Poisson bracket is given by
\[
\{F,G\}:=p\,\big(
\partial_p\,F\,\partial_x\,G-
\partial_x\,F\,\partial_p\,G\big)
,
\]
and the functions $S_a$ satisfy
\begin{equation}\label{wckp1}
\d\, S_a=m_a\,\d \,log\,z_a+\log\,p\,\d\,x-\sum_{a\neq 1,a_0}\,\log(p-q_a)\,\d\,t_{0a}+\sum_{a=1}^N\sum_{n\geq 1}\,\Omega_{na}\,\d\,t_{na}
.
\end{equation}

\subsection*{Appendix B: Proof of Proposition \eqref{tec.pro}}
The proof of  Proposition \eqref{tec.pro} requires the following  Lemma.
  \begin{lemma}\label{como actua T}\begin{enumerate}
     \item Given $T=\sum_{j\in\Z}c_j\Tp_a^j\in\T_a$, $a\neq 1$, then
      \begin{align}\label{TW1}
     & \left\{
      \begin{aligned}
             T(E_{11}W)&=T_>(W_{1k})E_{1k}
       +T_\leq(W_{11})E_{1}+\g_-W_0,\\
       T(E_{11}\bar W)&=\g_+\bar W_0,
        \end{aligned}
               \right.&a&=k\neq 1,\\%
     \label{TW2}&  \left\{\begin{aligned}
        T(E_{11}W)&=T_\leq(W_{11})E_{11}+
       \g_-W_0,\\
       T(E_{11}\bar W)&=T_>(\bar W_{1k})E_{1k}+\g_+\bar W_0,\\
       \end{aligned}
\right.&a&=\bar k,
     \end{align}
      \item Given $T=\sum_{j\in\Z}c_j\Tp_l^j\in\T_1$,  then
      \begin{align}\label{TW1l}
     & \left\{
      \begin{aligned}
             T(E_{11}W)&=T_\geq(W_{11})E_{11}
       +T_<(W_{1l_0})E_{1l_0}+\g_-W_0,\\
       T(E_{11}\bar W)&=\g_+\bar W_0,
        \end{aligned}
               \right.&a_0&=l_0\neq 1,\\%
    \label{TW2l} &  \left\{\begin{aligned}
        T(E_{11}W)&=T_\geq(W_{11})E_{11}+
       \g_-W_0,\\
       T(E_{11}\bar W)&=T_<(\bar W_{1l_0})E_{1l_0}+\g_+\bar W_0,\\
       \end{aligned}
\right.&a_0&=\bar {l_0},
     \end{align}
     \item Given $T=\sum_{j\in\Z}c_j(\bar\Tp_a^j-1)+c_0\in\bar \T_a$, $a\neq \bar 1,1$, then
      \begin{align}\label{barTW1}
       &  \left\{\begin{aligned}
        T(E_{11}W)&=T_>(W_{1k})E_{1k}+c_0W_{11}E_{11}+
       \g_-W_0,\\
       T(E_{11}\bar W)&=T_<(\bar W_{11})E_{11}+\g_+\bar W_0,
       \end{aligned}
\right.&a&= k\neq 1,\\\label{barTW2}
     & \left\{
      \begin{aligned}
             T(E_{11}W)&=c_0W_{11}E_{11}+\g_-W_0,\\
       T(E_{11}\bar W)&=T_>(\bar W_{lk})E_{lk}
       +T_<(\bar W_{11})E_{11}+\g_+\bar W_0,
        \end{aligned}
               \right.&a&=\bar k\neq \bar 1,
     \end{align}
      \item Given $T=\sum_{j\in\Z}c_j\bar\Tp_1^j\in\bar\T_1$,  then
      \begin{align}\label{barTWll}
        \left\{
      \begin{aligned}
             T(E_{11}W)&=T_\geq (W_{11})E_{11}+\g_-W_0,\\
       T(E_{11}\bar W)&=T_<(\bar W_{11})E_{11}+\g_+\bar W_0,
        \end{aligned}
               \right.
      \end{align}
      \item
       Given $T=\sum_{j\in\Z}c_j(\bar\Tp_{\bar 1}^j-1)+c_0\in\bar\T_{\bar 1}$, with $a_0\neq 1$,  then
      \begin{align}\label{barTW1l}
     & \left\{
      \begin{aligned}
             T(E_{11}W)&=T_<(W_{1l_0})E_{1l_0}+c_0W_{ll}E_{11}
     +\g_-W_0,\\
       T(E_{11}\bar W)&=T_>(\bar W_{11})+\g_+\bar W_0,
        \end{aligned}
               \right.&a_0&=l_0\neq 1,\\%
    \label{barTW2l} &  \left\{\begin{aligned}
        T(E_{11}W)&=c_0W_{11}E_{11}+
       \g_-W_0,\\
       T(E_{11}\bar W)&=T_>(\bar W_{11})E_{11}+T_<(\bar W_{1l_0})E_{1l_0}+\g_+\bar W_0,
       \end{aligned}
\right.&a_0&=\bar {l_0},
     \end{align}

     \item Given  $T=\sum_{j\in\Z}c_j\bar\Tp_{\bar 1}^j\in\bar\T_{\bar 1}$ with $a_0=1$, then
      \begin{align}\label{barTWll2}
        \left\{
      \begin{aligned}
             T(E_{11}W)&=T_\leq (W_{11})E_{11}+\g_-W_0,\\
       T(E_{11}\bar W)&=T_>(\bar W_{11})E_{11}+\g_+\bar W_0.
        \end{aligned}
               \right.
      \end{align}
                 \end{enumerate}
 \end{lemma}
\begin{proof}
    We only prove 1) since the others relations are proven similarly. From \eqref{TW1} observe that
    \begin{align*}
      \Tp_k^j(E_{11}W)&=\Tp_k^j(E_{11}S)\Tp_k^j(W_0)=\Tp_k^j(E_{11}S)(E_{kk}\Lambda^j+
      \I_N-E_{kk}-E_{11}+E_{11}
      \Lambda^{-j})W_0\\
      &=(\Tp_k^j(S_{1k})E_{1k}\Lambda^j+E_{11}\Tp_k^j(S)(\I_N-E_{kk}-E_{11})+\Tp_k^j(S_{11})
      E_{11}\Lambda^{-j})W_0,\\
      \Tp_k^j(E_{11}\bar W)&=\Tp_k^j(E_{11}\bar S)\bar W_0
    \end{align*}
    and therefore
    \begin{align*}
      \Tp_k^j(E_{11}W)E_{k'k'}&\in\g_-W_0, & k'\neq k,1,\\
        \Tp_k^j(E_{11}W)E_{kk}=\Tp_k^j(S_{1k})E_{1k}\Lambda^jW_0&\in\g_-W_0 & \text{ if $j\leq 0$},\\
        \Tp_k^j(E_{11}W)E_{11}=\Tp_k^j(S_{11})E_{11}\Lambda^{-j}W_0&\in\g_-W_0& \text{ if $j>0$},\\
         \Tp_k^j(E_{11}\bar W)&\in\g_+\bar W_0.
    \end{align*}
    Now, we check \eqref{TW2}. Notice that
     \begin{align*}
      \Tp_{\bar k}^j(E_{11}W)&=\Tp_{\bar k}^j(E_{11}S)\Tp_k^j(W_0)=\Tp_{\bar k}^j(E_{11}S)(\I_N-E_{11}+E_{11}\Lambda^{-j})W_0\\
      &=(E_{11}\Tp_{\bar k}^j(S)(\I_N-E_{11})+\Tp_{\bar k}^j(S_{11})E_{11}\Lambda^{-j})W_0,\\
      \Tp_{\bar k}^j(E_{11}\bar W)&=\Tp_{\bar k}^j(E_{11}\bar S)\Tp_{\bar k}^j(\bar W_0)=\Tp_{\bar k}^j(E_{11}\bar S)(E_{kk}\Lambda^{-j}+\I_N-E_{kk})\bar W_0\\
      &=(E_{11}\Tp_{\bar k}^j(\bar S)(\I_N-E_{kk})+\Tp_{\bar k}^j(\bar S_{1k})E_{1k}\Lambda^{-j})\bar W_0,
    \end{align*}
    and therefore
    \begin{align*}
      \Tp_{\bar k}^j(E_{11}W)E_{k'k'}&\in\g_-W_0, & k'\neq 1,\\
        \Tp_{\bar k}^j(E_{11}W)E_{ll}=\Tp_{\bar k}^j(S_{ll})E_{11}\Lambda^{-j}W_0&\in\g_-W_0, & \text{ if $j> 0$},\\
        \Tp_{\bar k}^j(E_{1}\bar W)E_{k'k'}&\in\g_+\bar W_0, &k'\neq k,\\
         \Tp_{\bar k}^j(E_{11}\bar W)E_{kk}=\Tp_{\bar k}^j(\bar S_{1k})\Lambda^{-j}\bar W_0&\in\g_+\bar W_0, & \text{ if $j\leq 0$},
    \end{align*}
\end{proof}

\begin{proof}[Proof of Proposition \ref{tec.pro}] The  proof of these results  relies on the previous Lemma \ref{como actua T} and Propositions
\ref{gd}, \ref{pro: shift orlov}. Let us go into details. We first
consider \eqref{F1}. From \eqref{TW1} we find for $k\neq 1$
\begin{align*}
  F(\M_k,\L_k)_>(E_{11}W)=F(\M_k,\L_k)(W_{1k})E_{1k}+\g_-W_0,
\end{align*}
so that, as we prove in Proposition \ref{pro: shift orlov}, we
deduce
\begin{align*}
    F(\M_k,\L_k)_>(E_{11}W)=E_{11}(F(M,L)C_{kk})W+\g_-W_0=E_{11}(F(M,L)C_{kk})_+W+\g_-W_0.
\end{align*}
Therefore,
\begin{align*}
  R:=F(\M_k,\L_k)_>(E_{11}W)-E_{11}(F(M,L)C_{kk})_+W\in\g_-W_0
\end{align*}
and from \eqref{facW} and \eqref{TW1} we get
\begin{align*}
  Rg=F(\M_k,\L_k)_>(E_{11}\bar W)-E_{11}(F(M,L)C_{kk})_+\bar W\in\g_+\bar W_0.
\end{align*}
so that Proposition \ref{gd} implies the first formula en \eqref{F1}.
Now, from \eqref{barTW1} we get for $k\neq 1$
\begin{align*}
  F(\bar \M_k,\bar\L_k)_>(E_{11}W)=F(\bar
  \M_k,\bar\L_k)(W_{1k})E_{1k}+\g_-W_0.
\end{align*}
Hence Proposition \ref{pro: shift orlov} ensures that
\begin{align*}
  R:= F(\bar \M_k,\bar\L_k)_>(E_{11}W)-E_{11}(F(M,L)C_{kk})_+W\in\g_-W_0
\end{align*}
and from \eqref{barTW1} we deduce
\begin{align*}
  Rg=F(\M_k,\L_k)_>(E_{11}\bar W)-E_{11}(F(M,L)C_{kk})_+\bar W\in\g_+\bar W_0.
\end{align*}
In this way,  Proposition \ref{gd} leads to the last formula in \eqref{F1}.
The proof of \eqref{F3} follows similarly.
\end{proof}

\section*{Acknowledgements}
The authors wish to thank the Spanish Ministerio de Ciencia e
Innovaci\'{o}n, research project FIS2008-00200,
and  acknowledge the support received from the European Science
Foundation (ESF) and the activity entitled \emph{Methods of Integrable
Systems, Geometry, Applied Mathematics} (MISGAM). MM wish to thank Prof. van Moerbeke and Prof. Dubrovin for
their warm hospitality, acknowledge
 economical support from MISGAM and  SISSA  and
 reckons different conversations with P. van Moerbeke,  T. Grava, G. Carlet and M. Caffasso. MM also acknowledges
 to Prof. Liu for his invitation  to visit the China Mining and Technology University at Beijing.

\end{document}